\documentclass{aastex631}

\usepackage{amsthm,amsmath,amssymb}
\usepackage{lipsum}
\usepackage{float}
\usepackage{soul}

\begin{document}

\title{BSN-III: The First Multiband Photometric Study on the Eight Total Eclipse Contact Binary Stars}

\author{Atila Poro}
\altaffiliation{atilaporo@bsnp.info, atila.poro@obspm.fr (AP)}
\affiliation{LUX, Observatoire de Paris, CNRS, PSL, 61 Avenue de l'Observatoire, 75014 Paris, France}
\affiliation{Astronomy Department of the Raderon AI Lab., BC., Burnaby, Canada}

\author{Kai Li}
\affiliation{School of Space Science and Technology, Institute of Space Sciences, Shandong University, Weihai, Shandong 264209, People's Republic of China}

\author{Raul Michel}
\altaffiliation{rmm@astro.unam.mx (RM)}
\affiliation{Instituto de Astronom\'ia, UNAM. A.P. 106, 22800 Ensenada, BC, M\'exico}

\author{Li-Heng Wang}
\affiliation{School of Space Science and Technology, Institute of Space Sciences, Shandong University, Weihai, Shandong 264209, People's Republic of China}

\author{Fahri Alicavus}
\affil{Çanakkale Onsekiz Mart University, Faculty of Sciences, Department of Physics, 17020, Çanakkale, Türkiye}
\affil{Çanakkale Onsekiz Mart University, Astrophysics Research Center and Ulupnar Observatory, 17020, Çanakkale,
Türkiye}

\author{Ghazal Alizadeh}
\affil{School of Aeronautics, Northwestern Polytechnical University, 710072, Xi'an, Shaanxi, People's Republic of China}

\author{Liliana Altamirano-D\'evora}
\affiliation{Instituto de Astronom\'ia, UNAM. A.P. 106, 22800 Ensenada, BC, M\'exico}
\affiliation{Facultad de Ingeniería, Arquitectura y Diseño (FIAD), Universidad Autónoma de Baja California (UABC), Ensenada, M\'exico}

\author{Francisco Javier Tamayo}
\affiliation{Facultad de Ciencias F\'{\i}sico-Matem\'aticas, UANL, 66451 San Nicol\'as de los Garza, NL, M\'exico}

\author{Hector Aceves}
\affiliation{Instituto de Astronom\'ia, UNAM. A.P. 106, 22800 Ensenada, BC, M\'exico}

\begin{abstract}
This study continues our in-depth investigation of total-eclipse W Ursae Majoris-type contact binaries by analyzing eight new systems, complementing our previous work. Multiband $BVR_cI_c$ photometric data were acquired through ground-based observations at an observatory in Mexico, from which new times of minima were determined. Our analysis of orbital period variations using the O-C method revealed that one system shows no long-term variation, four systems exhibit a secular decrease in their orbital periods, and two systems exhibit a secular increase, suggesting mass transfer between the components. Notably, one system displays a cyclic variation with an amplitude of 0.00865 days and a period of 10.49 years, which we attribute to the light travel time effect induced by a tertiary companion, possibly a brown dwarf. We modeled the light curves using the PHOEBE Python code. Six of the target systems required the inclusion of a cold starspot on one of the system's stars due to the asymmetry observed in the maxima of their light curves. Absolute parameters were estimated using the Gaia DR3 parallax method. Using the components' effective temperatures and masses, we classified five of the systems as W-subtype and three as A-subtype. The stellar evolution was illustrated through the mass-radius and mass-luminosity diagrams. Furthermore, we investigated the dynamical stability of two systems with extremely low mass ratios.
\end{abstract}


\section{Introduction}
\label{sec1}
Binary star systems are typically classified into three categories based on their configuration in the Roche potential (\citealt{1968ApJ...151.1123L}, \citealt{1959cbs..book.....K}): detached, semidetached, and contact binary systems. W Ursae Majoris (W UMa)-type contact binaries are characterized by their distinctive light curves, which exhibit nearly equal-depth minima, continuous brightness variations, and short orbital periods (\citealt{2014ApJS..212....4Q}). In these systems, both stellar components share a common convective envelope, resulting in nearly identical surface temperatures despite possible mass differences (\citealt{1968ApJ...151.1123L}, \citealt{1968ApJ...153..877L}). This shared envelope facilitates efficient thermal contact and enables the transfer of both mass and energy between the components.

Mass transfer and angular momentum loss in contact binaries are fundamental processes influencing the evolution of the system as a whole, including changes in the orbital period. Consequently, it is reasonable to anticipate correlations between the orbital period and various physical parameters of these systems. Although numerous studies have examined these relationships, a comprehensive and coherent understanding is still open to debate (e.g., \citealt{2021ApJS..254...10L}, \citealt{2024RAA....24a5002P}, \citealt{2025MNRAS.538.1427P}). These efforts highlight the intricate interplay of mass exchange, angular momentum loss, and thermal equilibrium that collectively shape the evolutionary pathways of contact binaries. In addition, the presence of a third body adds complexity to the evolution of the system as a whole, affecting the orbital dynamics as well as other system properties (\citealt{2023AA...678A..60K}, \citealt{2024NewA..10502112S}, \citealt{2024RAA....24d5018P}).

Most contact binary systems exhibit effective temperatures in the range of approximately 3500 K to 7200 K. According to the study by \cite{2021ApJS..254...10L}, systems with both an orbital period longer than 0.5 days and an effective temperature around 7000 K are not classified as W UMa-type binaries. Determining the surface temperature, along with the masses of the component stars, allows for the classification of a contact binary into A-type or W-type subtypes (\citealt{binnendijk1970}). In A-type contact binaries, the more massive component is also the hotter star, whereas in W-type binaries, the less massive component has a higher effective temperature.

Despite decades of investigations, several significant issues remain unresolved in the study of contact binaries. These include the orbital period cut-off \citealt{2020MNRAS.497.3493Z}, the stability of systems with very low mass ratios \citealt{2022AJ....164..202L}, \citealt{2024MNRAS.527....1W}, the accurate determination of mass ratios from photometric light curves \citealt{2023ApJ...958...84K}, and the empirical relationships between parameters such as orbital period and mass ratio, or mass and luminosity \citealt{2024RAA....24a5002P}. In addition to these challenges, several other theoretical aspects remain poorly understood, particularly the mechanism of energy transfer between the stellar components, which is fundamental to the structure and long-term evolution of contact systems (\citealt{1968ApJ...151.1123L}, \citealt{2023AA...672A.175F}). These issues all require further investigation. Exploring additional examples of these binary star types, particularly those that have not yet been investigated, will significantly help enhance our understanding of contact systems.

This study presents a detailed photometric analysis of eight W UMa-type contact binaries undergoing total eclipses. Also, this work continues the investigation initiated by \cite{2025MNRAS.537.3160P} and \cite{2025.AJ.P}, presenting new observations and an in-depth analysis of more W UMa-type contact binary systems in the BSN\footnote{\url{https://bsnp.info/}} project. The paper is structured as follows: Section 2 outlines the basic characteristics of the target systems. Section 3 describes the observation and data reduction processes. Section 4 focuses on the analysis of orbital period variations, and Section 5 presents the light curve modeling results. Section 6 provides estimations of the absolute parameters, while Section 7 discusses the results and presents the conclusions.

\vspace{0.6cm}
\section{Target Systems}
\label{sec2}
We have analyzed eight eclipsing binary stars, including BU Tri, CRTS J170839.8+122530 (hereinafter J1708) CRTS J115758.8+331718 (hereinafter J1157), CRTS J123651.3-070549 (hereinafter J1236), CRTS J164801.9+451118 (hereinafter J1648), CRTS J233315.9+355134 (hereinafter J2333), V1232 Her, V1487 Her, ZTF J161614.70+162306.9 (hereinafter Z1616). These target contact binary systems analyzed in this study were selected randomly based on two main criteria. First, the systems had not been studied in detail previously. Second, multiband photometric data are available for these targets in the BSN project database, providing sufficient observational coverage for accurate analysis. Table \ref{systemsinfo} presents specifications for the target systems based on the Gaia DR3 database (\citealt{2023AA...674A..33G}), and standard notation for other quantities is used. The general properties of the target systems are summarized below:

$\bullet$ BU Tri: This system was discovered by \cite{2007OEJV...58....1L} as an eclipsing binary in the field of RV Tri. BU Tri is recognized in various catalogs—including the All-Sky Automated Survey for Supernovae (ASAS-SN), the Variable Star Index (VSX), and the Zwicky Transient Facility (ZTF, \citealt{2023AA...675A.195S})—as a contact binary system. The orbital period of BU Tri is consistent across most catalogs, agreeing to the fourth decimal place. The VSX database reports an orbital period of 0.295562 days and a maximum apparent magnitude of $14.400^{mag.}$ for the system.

$\bullet$ J1157: This eclipsing binary system was identified in the Catalina Surveys Data Release 1 (CSDR1, \citealt{2014ApJS..213....9D}). Both the ASAS-SN and ZTF catalogs of periodic variable stars report an orbital period of 0.3412135 days for the system. The VSX database gives a maximum apparent magnitude of $14.740^{\mathrm{mag}}$ for J1157. This system is the hottest target in this study (Table \ref{systemsinfo}).

$\bullet$ J1236: This eclipsing binary system was discovered in the CSDR1 catalog (\citealt{2014ApJS..213....9D}). This system is also known as a contact binary system in other catalogs, such as ZTF which reports an orbital period of 0.2996344 days. The VSX database reports a maximum apparent magnitude of $14.350^{mag.}$ for J1236, with a variability amplitude of $0.25^{mag.}$. However, the amplitude reported in the ASAS-SN catalog is $0.32^{mag.}$.

$\bullet$ J1648: This system was discovered in the Trans-atlantic Exoplanet Survey (TrES, \citealt{2004ApJ...613L.153A}) project as an eclipsing binary. The ASAS-SN, ZTF, and VSX catalogs introduce this system as a contact binary with an orbital period of 0.31372 days. According to the VSX database, J1648 has a maximum apparent magnitude of $14.430^{mag.}$.

$\bullet$ J2333: The discovery of this eclipsing binary system was first reported in CSDR1 (\citealt{2014ApJS..213....9D}). J2333 is classified as a contact binary system in the CSDR1, ZTF, ASAS-SN, and Asteroid Terrestrial-impact Last Alert System (ATLAS, \citealt{2018AJ....156..241H}) catalogs. The maximum apparent magnitude of this system is $14.390^{mag.}$, and its orbital period is 0.294013, as reported in the VSX database.

$\bullet$ V1232 Her: This binary system was discovered by the Robotic Optical Transient Search Experiment I (ROTSE-I) telescope, which presented the first results of a search for periodic variable stars (\citealt{2000AJ....119.1901A}). The orbital period of this system is listed as 0.2679 days in the ASAS-SN, ZTF, and VSX catalogs. Additionally, the VSX catalog provides a maximum magnitude of $14.450^{mag.}$ for V1232 Her.

$\bullet$ V1487 Her: The eclipsing binary system was first discovered from the CSDR1 (\citealt{2014ApJS..213....9D}), which provided extensive time-series photometry for variable star detection. V1487 Her is listed with an orbital period of 0.2244967 days in both the ASAS-SN and ZTF catalogs. Its maximum magnitude is $15.210^{mag.}$ in the VSX database. This system has the lowest effective temperature reported by Gaia DR3 among all the targets included in this study.

$\bullet$ Z1616: This binary system was discovered by ZTF (\citealt{2020ApJS..249...18C}). Z1616, located in the Hercules constellation, has been classified as a contact system in well-known catalogs such as ASAS-SN, VSX, and ZTF. The orbital period of this system is reported 0.2713132 days in the ZTF catalog. Z1616 is a faint system, with a maximum apparent magnitude of $15.307^{mag.}$ reported in the VSX database.

\begin{table*}
\renewcommand\arraystretch{1.2}
\caption{Specifications of the target systems from the Gaia DR3.}
\centering
\begin{center}
\footnotesize
\begin{tabular}{c c c c c c c}
\hline
System & RA$.^\circ$(J2000) & Dec$.^\circ$(J2000) & $d$(pc) & RUWE & $T_{Gaia}$(K) & $V-R$(mag.)\\
\hline
BU Tri	&	33.256231	&	37.057179	&	1626(98)	&	1.004	&	5784	& 0.402\\
CRTS J115758.8+331718 (J1157)	&	179.495280	&	33.288289	&	1931(118)	&	1.124	&	6973	& 0.203\\
CRTS J123651.3-070549 (J1236)	&	189.214048	&	-7.097267	&	923(19)	&	1.032	&	5619	& 0.371\\
CRTS J164801.9+451118 (J1648)	&	252.008279	&	45.188201	&	904(12)	&	1.036	&	5785	& 0.344\\
CRTS J233315.9+355134 (J2333)	&	353.316143	&	35.859840	&	841(14)	&	1.006	&	5040	& 0.461\\
V1232 Her	&	254.669272	&	37.771728	&	679(8)	&	1.151	&	5301	& 0.426\\
V1487 Her	&	254.394284	&	27.802909	&	737(14)	&	1.045	&	4911	& 0.491\\
ZTF J161614.70+162306.9 (Z1616)	&	244.061303	&	16.385240	&	650(10)	&	0.990	&	5182	& 0.342\\
\hline
\end{tabular}
\end{center}
\label{systemsinfo}
\end{table*}

\vspace{0.6cm}
\section{Observation and Data Reduction}
\label{sec3}
Observations of the eight binary systems were performed at the San Pedro Mártir (SPM) Observatory in México, situated at $115^\circ$  $27^{'}$ $49^{''}$ West and $31^\circ$ $02^{'}$ $39^{''}$ North, at an elevation of 2830 meters above sea level.

These observations were conducted using two Ritchey-Chrétien telescopes. The 0.84-meter telescope with an $f/15$ focal ratio was paired with the Mexman filter wheel and the Marconi 5 CCD detector, an e2v CCD231-42 featuring $15 \times 15 \, \mu\mathrm{m}^2$ pixels, a gain of $2.2 \, e^- /\mathrm{ADU}$, and a readout noise of $3.6 \, e^-$. The 1.5-meter telescope utilized the RUCA filter wheel and the Spectral Instruments 1 detector, which includes an e2v CCD42-40 with $13.5 \times 13.5 \, \mu\mathrm{m}^2$ pixels, a gain of $1.39 \, e^- /\mathrm{ADU}$, and a readout noise of $3.49 \, e^-$. Observations were carried out using standard $B$, $V$, $R_c$, and $I_c$ filters.

The photometric data were processed using IRAF software routines, following the procedures outlined by \cite{1986SPIE..627..733T}. Standard data reduction steps, including bias subtraction and flat-field correction, were applied.

Table \ref{observations} outlines the main observational parameters for each target, such as observation dates, filters utilized, and exposure times. Additionally, Table \ref{stars} presents the coordinates of the comparison and check stars identified during the observation and data reduction processes. These stars were essential for ensuring the accuracy and stability of our photometric measurements. Specifically, the comparison stars were used as reference points to calibrate the brightness of the target binaries, while the check stars served to verify the constancy of the comparison stars throughout the observing sessions. This approach helped to minimize systematic errors and improve the reliability of the final light curves. The information in Table \ref{stars} is from Gaia DR3 \citep{2023AA...674A..33G}.

\begin{table*}
\renewcommand\arraystretch{1.2}
\caption{Specifications of the ground-based observations.}
\centering
\begin{center}
\footnotesize
\begin{tabular}{c c c c}
\hline
System & Observation(s) Date & Filter & Exposure time(s)\\
\hline
BU Tri	&	 2024 (October 6) 	&	 $BVR_cI_c$ 	&	 $B(90)$, $V(50)$, $R_c(35)$, $I_c(30)$ 	\\
J1157	&	 2024 (April 7) 	&	 $BVR_cI_c$ 	&	 $B(80)$, $V(40)$, $R_c(30)$, $I_c(30)$ 	\\
J1236	&	 2024 (April 8, April 12) 	&	 $BVR_cI_c$ 	&	 $B(40)$, $V(20)$, $R_c(15)$, $I_c(15)$ 	\\
J1648	&	 2024 (May 23) 	&	 $BVR_cI_c$ 	&	 $B(70)$, $V(50)$, $R_c(35)$, $I_c(30)$ 	\\
J2333	&	 2024 (September 13) 	&	 $BVR_cI_c$ 	&	 $B(90)$, $V(50)$, $R_c(30)$, $I_c(25)$ 	\\
V1232 Her	&	 2024 (May 31) 	&	 $BVR_cI_c$ 	&	 $B(70)$, $V(50)$, $R_c(35)$, $I_c(30)$ 	\\
V1487 Her	&	 2024 (May 29) 	&	 $BVR_cI_c$ 	&	 $B(70)$, $V(50)$, $R_c(35)$, $I_c(30)$ 	\\
Z1616	&	 2024 (May 17) 	&	 $BVR_cI_c$ 	&	 $B(60)$, $V(30)$, $R_c(20)$, $I_c(20)$ 	\\
\hline
\end{tabular}
\end{center}
\label{observations}
\end{table*}

\begin{table*}
\renewcommand\arraystretch{1.2}
\caption{List the comparisons and check stars in the ground-based observations.}
\centering
\begin{center}
\footnotesize
\begin{tabular}{c c c c c c}
\hline
System & Star Type & Star Name & RA$.^\circ$(J2000) & DEC$.^\circ$(J2000) & $V-R$(mag.)\\
\hline
BU Tri	&	Comparison	&	Gaia DR3 331351680302769536 & 33.349451	&	37.154424	&	0.425	\\
BU Tri	&	Check	&	Gaia DR3 331351714662506496 & 33.377602	&	37.157921	&	0.383	\\
J1157	&	Comparison	&	Gaia DR3 4027982100132827904 & 179.555646	&	33.296308	&	0.302	\\
J1157	&	Check	&	Gaia DR3 4027619914130577536 & 179.435020	&	33.369127	&	0.371	\\
J1236	&	Comparison	&	Gaia DR3 3676452427253464064 & 189.225505	&	-7.096860	&	0.376	\\
J1236	&	Check	&	Gaia DR3 3676452491677347840 & 189.226342	&	-7.087546	&	0.334	\\
J1648	&	Comparison	&	Gaia DR3 1407379938730847872 & 252.065931	&	45.188669	&	0.321	\\
J1648	&	Check	&	Gaia DR3 1407392686194264448 & 251.963374	&	45.246438	&	0.409	\\
J2333	&	Comparison	&	Gaia DR3 1912581892195834368 & 353.371277	&	35.854498	&	0.457	\\
J2333	&	Check	&	Gaia DR3 1912581033202376832 & 353.376838	&	35.830811	&	0.431	\\
V1232 Her	&	Comparison	&	Gaia DR3 1351660228488885504 & 254.671414	&	37.692597	&	0.460	\\
V1232 Her	&	Check	&	Gaia DR3 1351671773360987136 & 254.583832	&	37.678340	&	0.444	\\
V1487 Her	&	Comparison	&	Gaia DR3 1307108360929012096 & 254.385649	&	27.814172	&	0.499	\\
V1487 Her	&	Check	&	Gaia DR3 1307109323001695872 & 254.315983	&	27.835062	&	0.422	\\
Z1616	&	Comparison	&	Gaia DR3 4465272683548078848 & 244.052929	&	16.385713	&	0.410	\\
Z1616	&	Check	&	Gaia DR3 4465272717907815168 & 244.071211	&	16.389799	&	0.450	\\
\hline
\end{tabular}
\end{center}
\label{stars}
\end{table*}

\vspace{0.6cm}
\section{Orbital Period Variations}
\label{sec4}
\begin{figure*}
\centering
\includegraphics[width=0.48\textwidth]{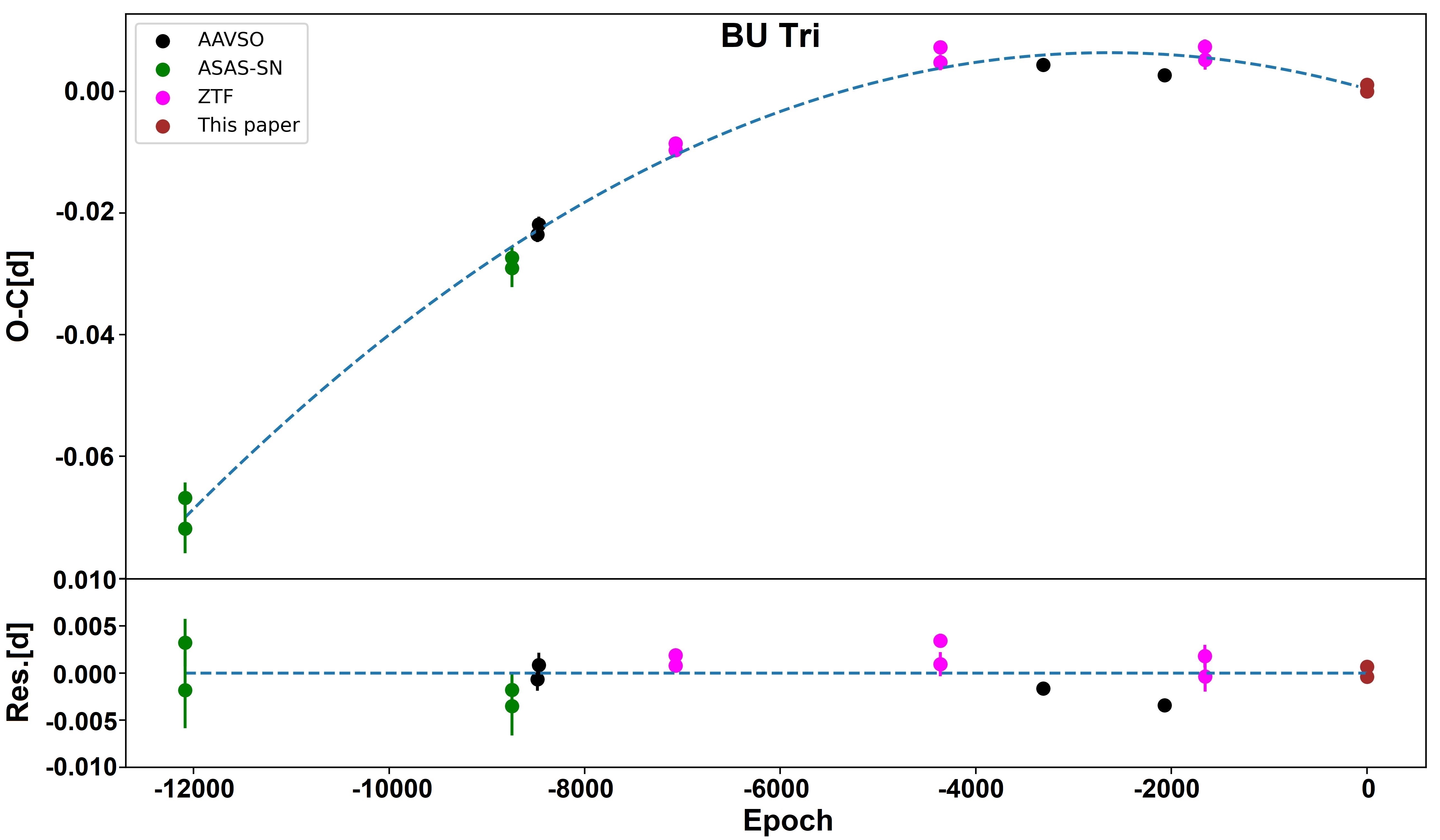}
\includegraphics[width=0.48\textwidth]{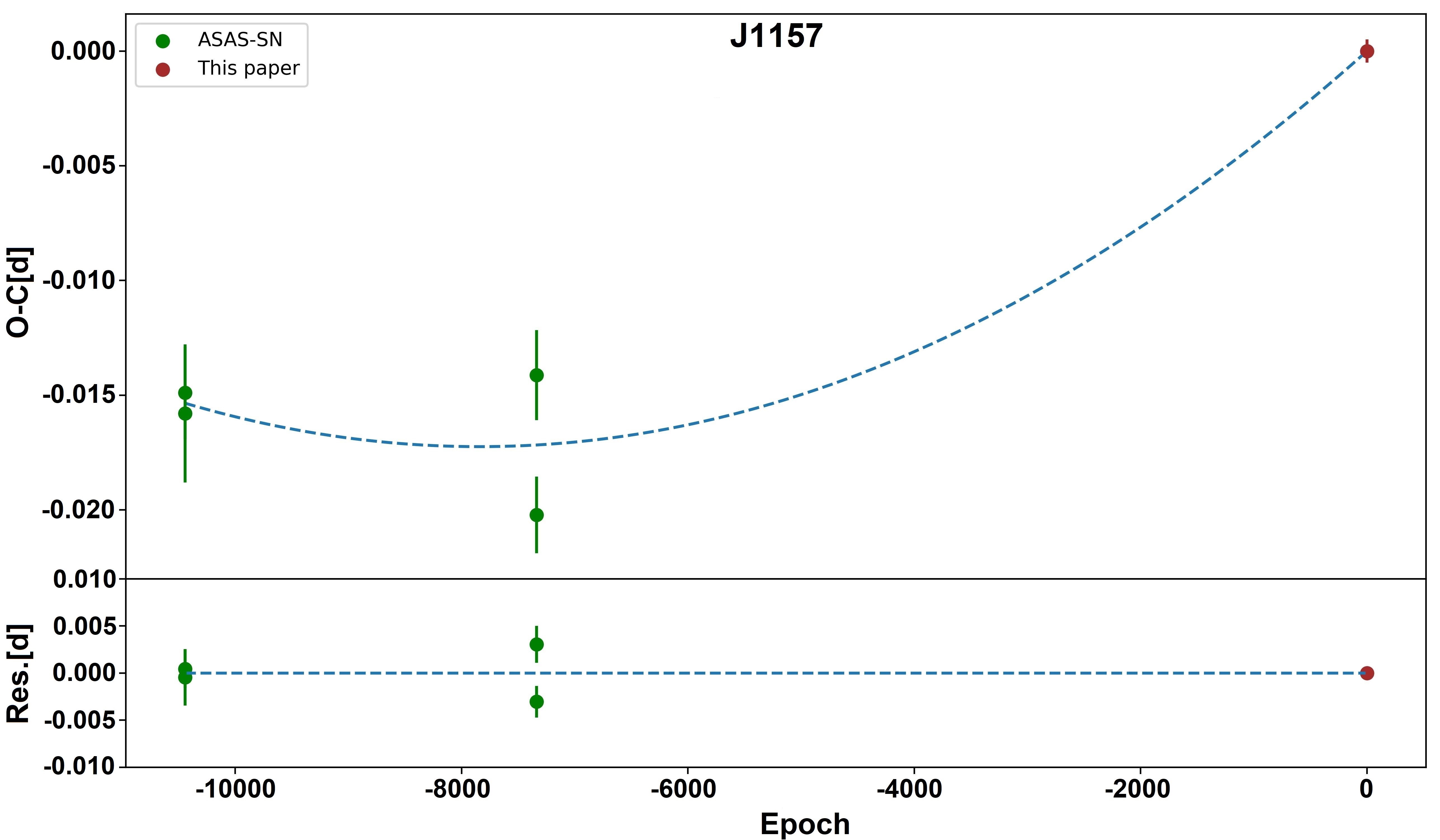}
\includegraphics[width=0.48\textwidth]{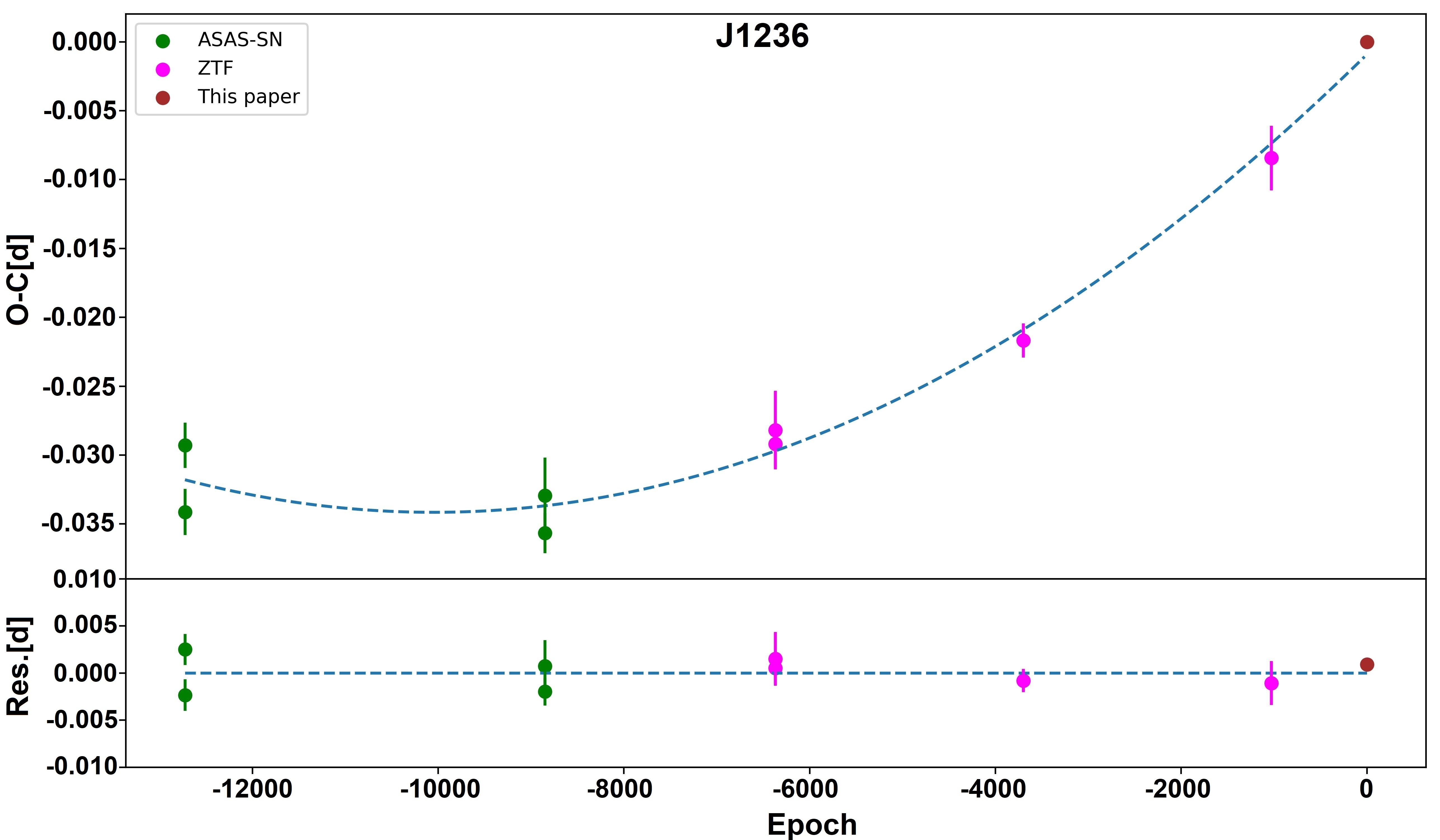}
\includegraphics[width=0.48\textwidth]{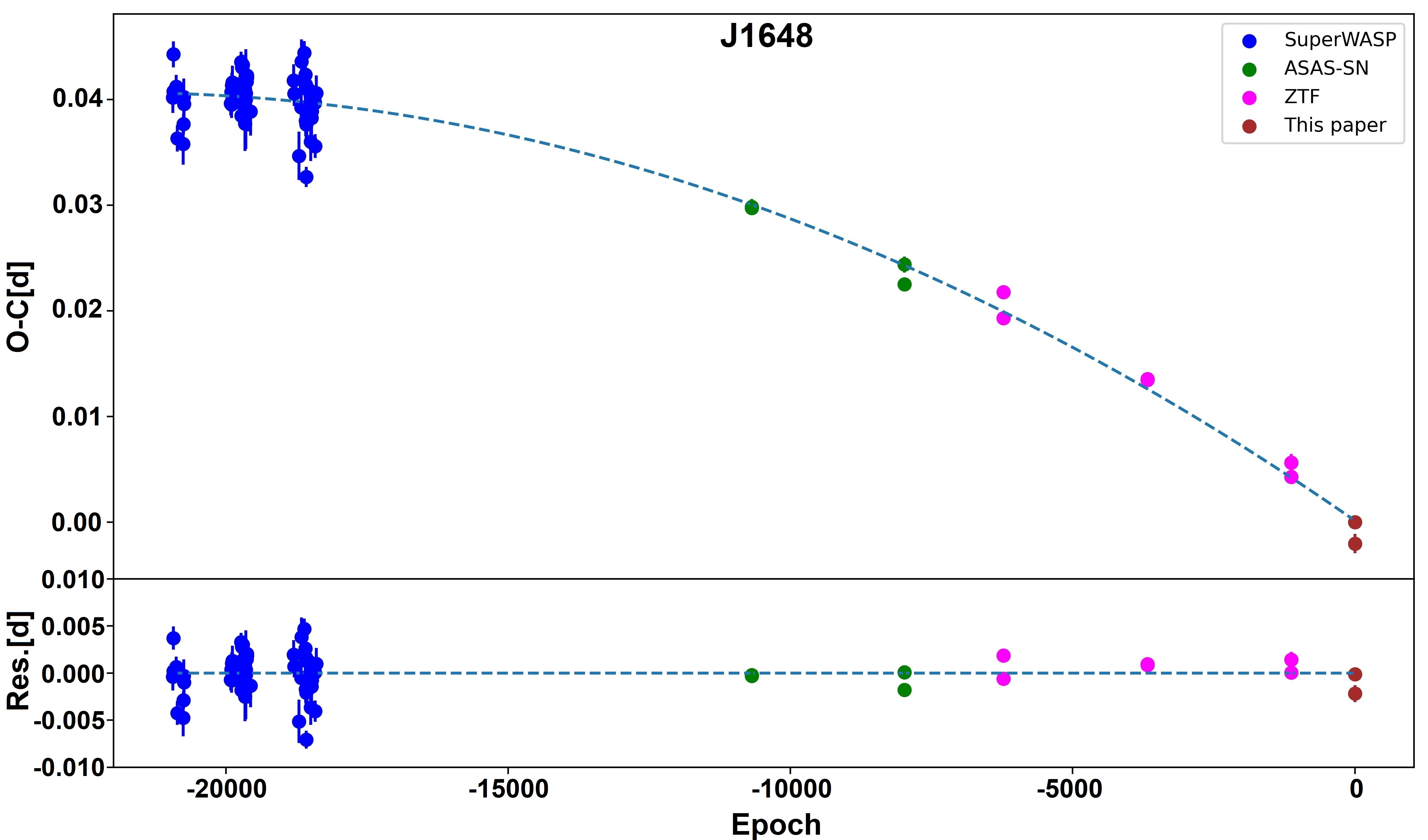}
\includegraphics[width=0.48\textwidth]{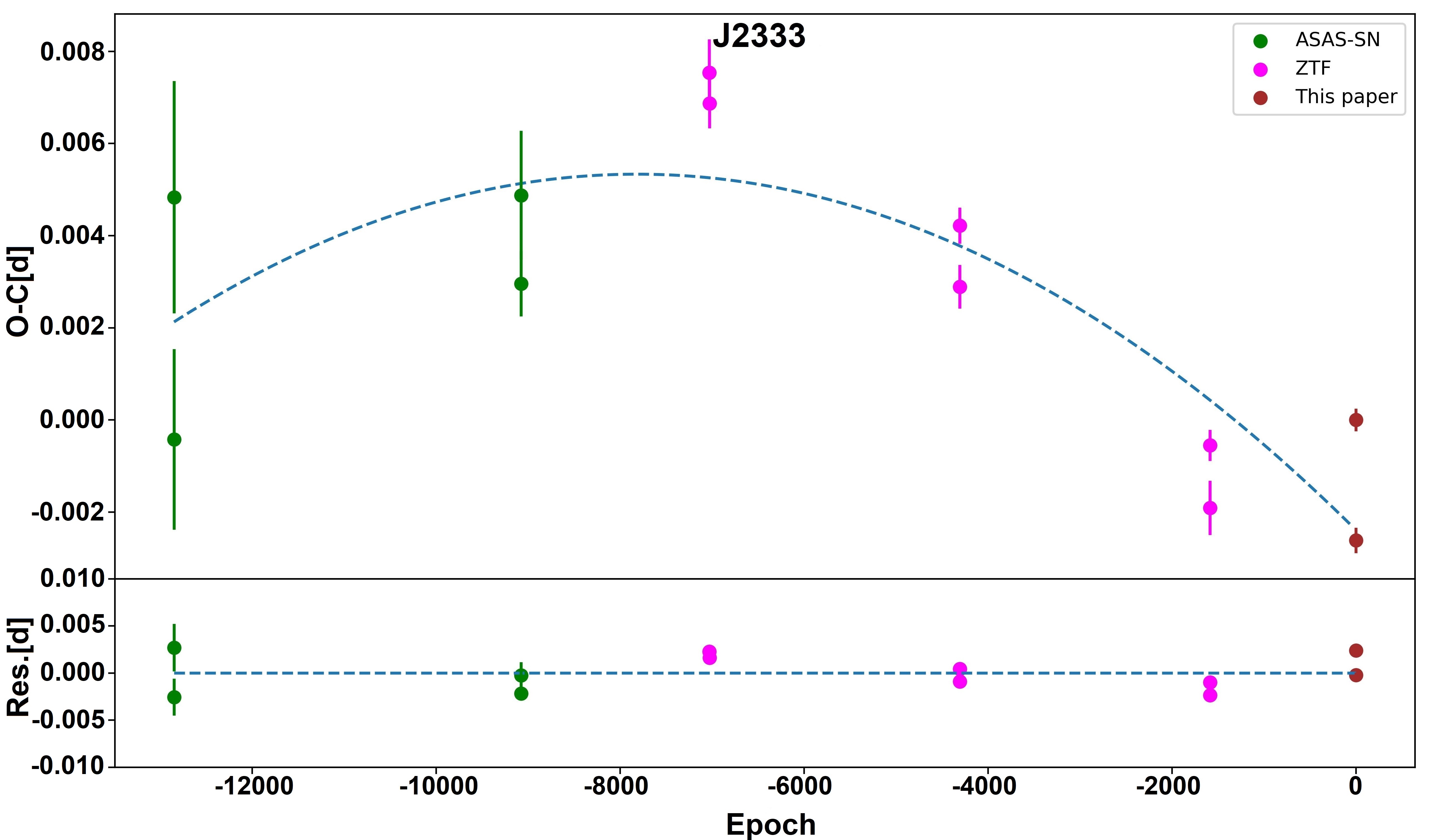}
\includegraphics[width=0.48\textwidth]{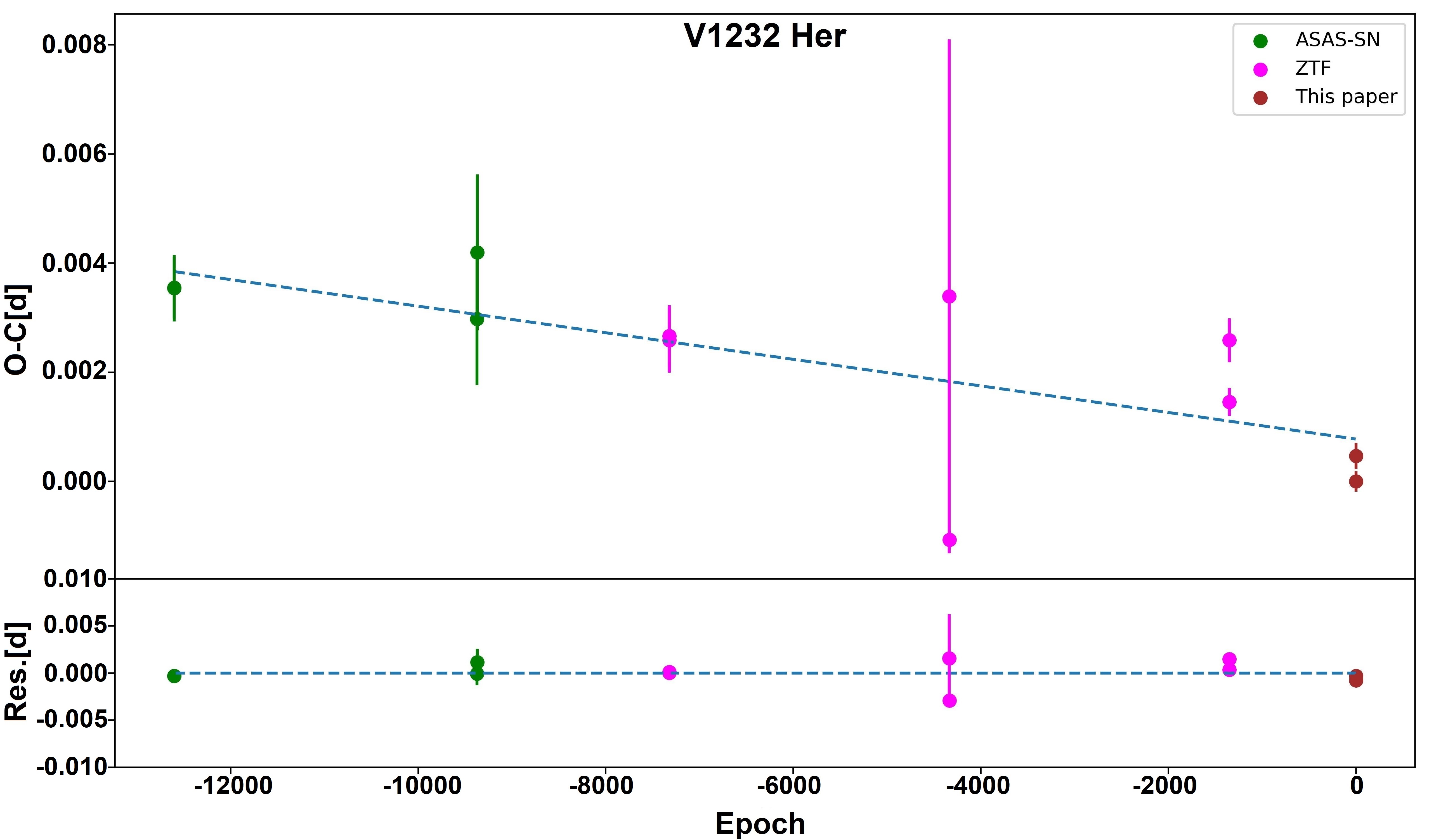}
\includegraphics[width=0.48\textwidth]{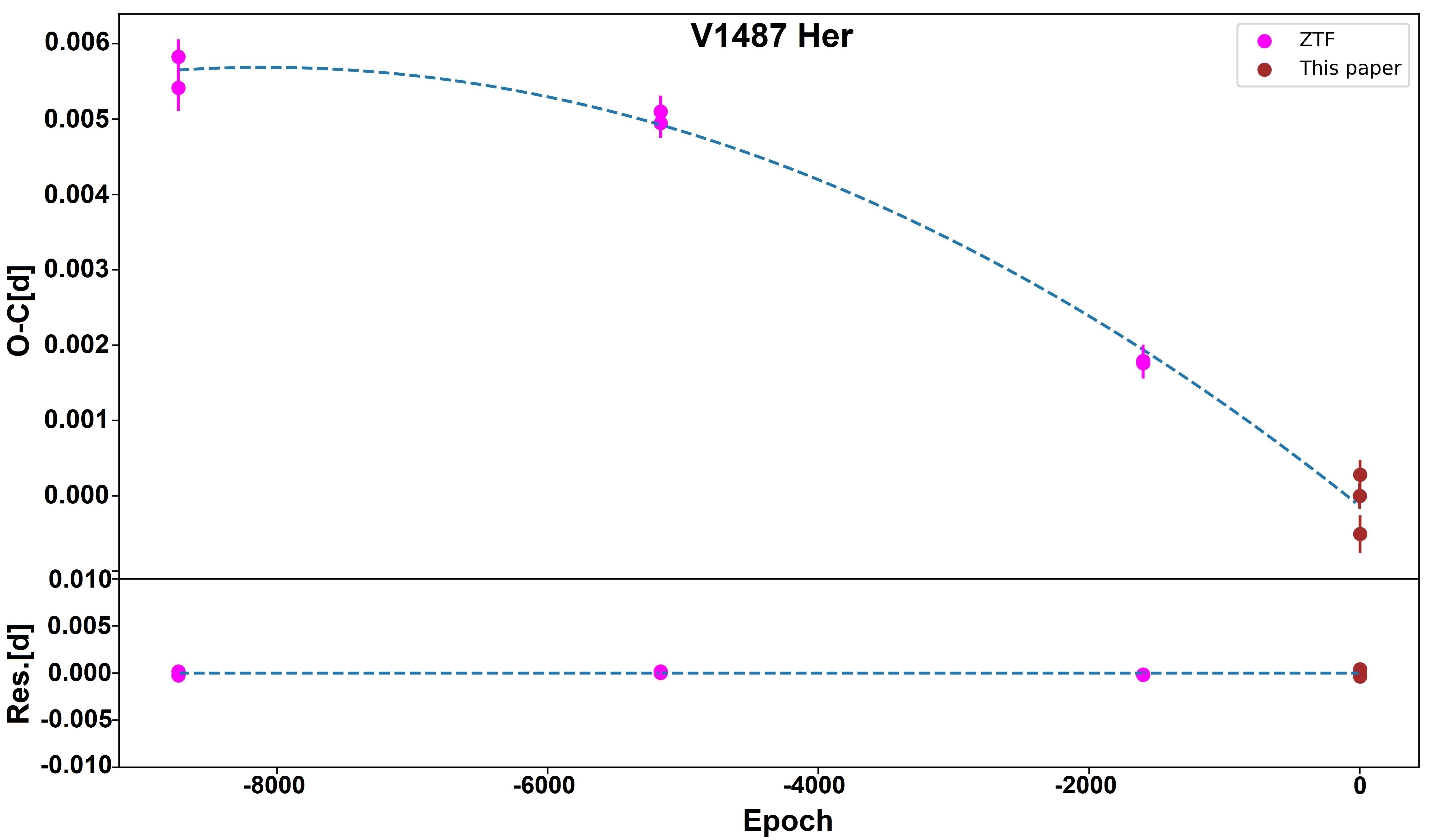}
\includegraphics[width=0.48\textwidth]{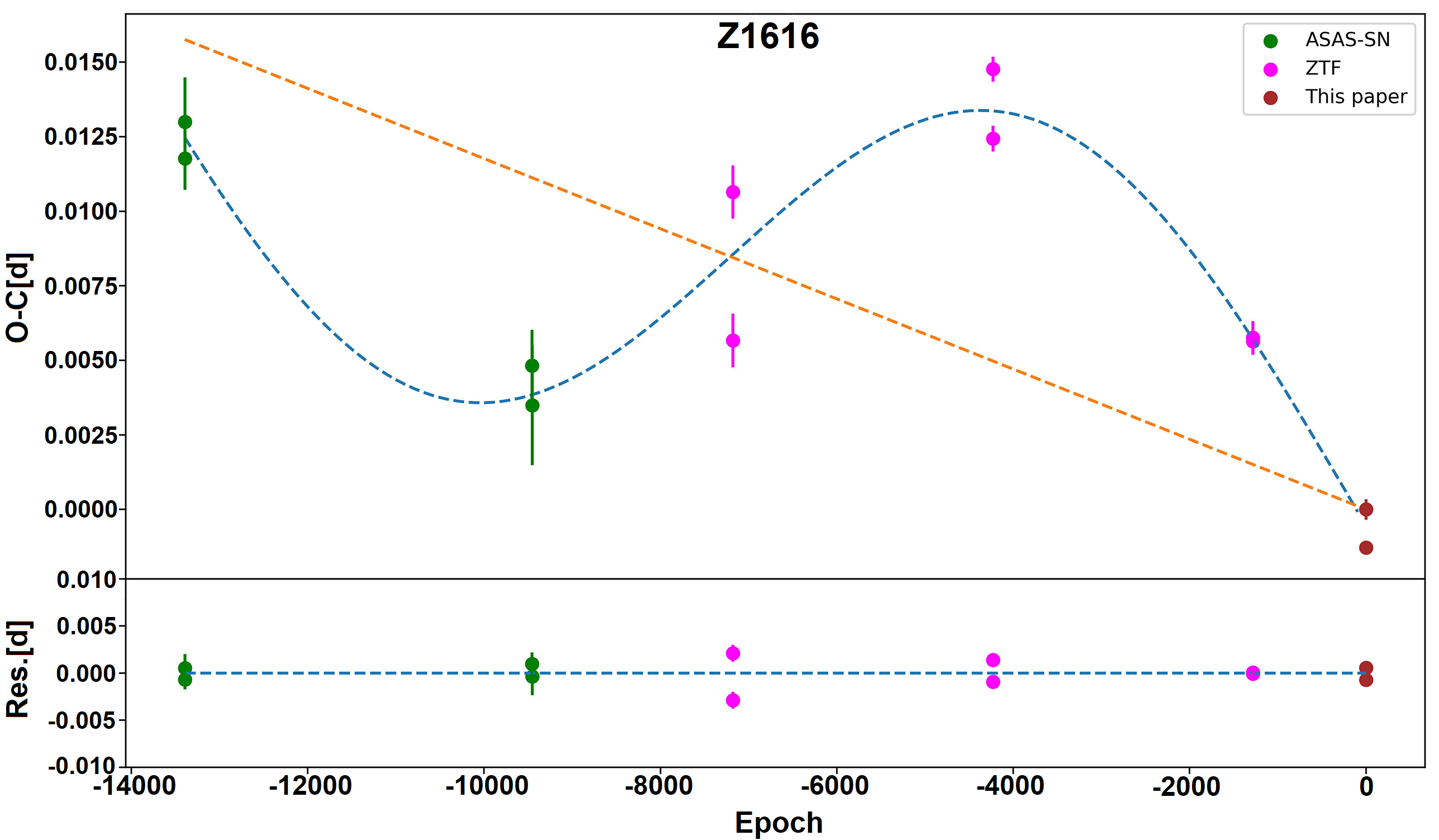}
\caption{The O-C diagrams of the eight targets, with residuals at the bottom}.
\label{O-CFigs}
\end{figure*}

The O-C (observed-minus-calculated) method is utilized to investigate the variations in the orbital periods of our eight binary star systems. In order to obtain as many eclipse times as possible, we used the photometric survey data from a variety of sources, including All-Sky Automated Survey for SuperNovae \cite[ASAS-SN;][]{shappee2014,jayasinghe2018}, the Zwicky Transient Facility \cite[ZTF;][]{ztf1,ztf2}, the Transiting Exoplanet Survey Satellite \cite[TESS;][]{tess}, Wide Angle Search for Planets \cite[SuperWASP;][]{wasp}, and American Association of Variable Star Observers (AAVSO). Regarding the data from AAVSO, SuperWASP, and TESS, we were able to directly calculate the eclipse times employing the method described by \cite{kw1956}. Conversely, for the data obtained from ASAS-SN and ZTF, we utilized the period shift technique introduced by \cite{Li_2020}. This involved first consolidating the discrete data points into a single period, before proceeding to compute the time of the eclipse minimum. The times of eclipse minima were subsequently converted from the Heliocentric Julian Date ($HJD$) to Barycentric Julian Date in Barycentric Dynamical Time ($BJD_{TDB}$) using the online transformation tool of \cite{Eastman2010}\footnote{\url{https://astroutils.astronomy.osu.edu/time/hjd2bjd.html}}. The eclipsing times extracted from our observations are listed in Table \ref{min}. The online machine-readable format is available for the extracted and collected minima times of the target binary systems. To detect orbital period variations, we computed O-C values using the following linear ephemeris,

\begin{equation}
\begin{aligned}
T=T_0+P\times E,
\end{aligned}
\end{equation}

\noindent where $T$ is the observed eclipse times, $T_0$ is the reference primary eclipse time listed in the second column of Table \ref{ephemeris}, and $P$ is the orbital period listed in the third column of Table \ref{ephemeris}, $E$ is the cycle number. The calculated O-C values are listed in Table \ref{min} and online machine-readable format are available. The corresponding O-C diagram is shown in Figure \ref{O-CFigs}. We found that six of our systems show secular trends. The following equation was used to fit their O-C diagrams,

\begin{equation}
\begin{aligned}
O-C=\Delta{T_0}+\Delta{P_0}\times E+\frac{\beta}{2}{E^2}.
\end{aligned}
\end{equation}

The derived parameters are presented in Table \ref{OCtab} (the mass transfer rate was calculated by Equation \ref{7} for fully conservative mass transfer without angular momentum loss), while the corrected new ephemerides are detailed in Table \ref{ephemeris}. One system (V1232 Her) shows no long-term variation, hence a linear fit was used to fit its O-C curve, and the fitted parameters and the corrected new ephemeris are shown in Tables \ref{OCtab} and \ref{ephemeris}. One system (Z1616) show cyclic variation, the following equation was used to fit its O-C curve,

\begin{equation}
\begin{aligned}
O-C=\Delta{T_0}+\Delta{P_0}\times E+A\sin(\frac{2\pi}{P_3}\times E+\varphi).
\end{aligned}
\end{equation}

The derived $\Delta{T_0}$ and $\Delta{P_0}$ and the corrected new ephemeris are also shown in Tables \ref{OCtab} and \ref{ephemeris}. The amplitude and the period of the cyclic variation are determined to be $A=0.00865\pm0.00784$ d and $P_3=10.49\pm5.35$ yr.
Based on the analysis performed, we found that four systems exhibit long-term decrease orbital period, two systems show long-term increase orbital period, and one system shows cyclic variation orbital period.

\renewcommand\arraystretch{1.2}
\begin{table*}
\caption{The times of minima extracted from our ground-based observations.}
\centering
\small
\begin{tabular}{c c c c c}
\hline
System & Min.($BJD_{TDB}$) & Error & Epoch & O-C\\ 
\hline
BU Tri	&	2460589.82864	&	0.00047	&	0.0	&	0.00000	\\
BU Tri	&	2460589.97749	&	0.00037	&	0.5	&	0.00107	\\
J1157	&	2460407.86990	&	0.00051	&	-0.5	&	0.00000	\\
J1236	&	2460408.79884	&	0.00017	&	0.0	&	0.00000	\\
J1648	&	2460453.69570	&	0.00056	&	0.0	&	0.00000	\\
J1648	&	2460453.85051	&	0.00091	&	0.5	&	-0.00204	\\
J2333	&	2460566.68979	&	0.00025	&	0.0	&	0.00000	\\
J2333	&	2460566.83418	&	0.00028	&	0.5	&	-0.00262	\\
V1232 Her	&	2460461.76706	&	0.00019	&	0.0	&	0.00000	\\
V1232 Her	&	2460461.90147	&	0.00024	&	0.5	&	0.00047	\\
V1487 Her	&	2460459.69951	&	0.00017	&	0.0	&	0.00000	\\
V1487 Her	&	2460459.81125	&	0.00026	&	0.5	&	-0.00051	\\
V1487 Her	&	2460459.92428	&	0.00020	&	1.0	&	0.00028	\\
Z1616	&	2460447.70198	&	0.00034	&	0.0	&	0.00000	\\
Z1616	&	2460447.83636	&	0.00019	&	0.5	&	-0.00128	\\
\hline
\end{tabular}
\label{min}
\end{table*}

\renewcommand\arraystretch{1.2}
\begin{table*}
\caption{\centering Reference and new ephemeris of the eight systems. The reference times of minimum ($t_0$) were obtained from our observations in this study.}
\centering
\small
\begin{tabular}{c|cc|cc}
\hline
System& \multicolumn{2}{c|}{Reference ephemeris}& \multicolumn{2}{c}{New ephemeris}\\ 
&$t_0(BJD_{TDB})$&Period(day)/Source& Corrected $t_0(BJD_{TDB})$&New Period(day)\\ 
\hline
BU Tri
&  2460589.8286(5)&0.2955620/ASAS-SN& 2460589.8290(26)&0.2955575(10)\\ 
J1157&  2460408.0405(5)&0.3412135/ASAS-SN& 2460408.0405(62)&0.3412179(26)\\ 
J1236&  2460408.7988(2)&0.2996346/ASAS-SN& 2460408.7979(29)&0.2996412(10)\\ 
J1648&  2460453.6957(6)&0.3137180/ASAS-SN& 2460453.6958(20)&0.3137143(5)\\ 
J2333&  2460566.6898(2)&0.2940043/ASAS-SN& 2460566.6874(25)&0.2940023(9)\\
 V1232 Her
& 2460461.7671(2)& 0.2679032/ASAS-SN& 2460461.7679(12)&0.2679030(2)\\
 V1487 Her
& 2460459.6995(2)& 0.2244968/VSX& 2460459.6994(3)&0.2244954(2)\\
 Z1616& 2460447.7020(3)& 0.2713132/ASAS-SN& 2460447.7020(154)& 0.2713120(22)\\
 \hline
\end{tabular}
\label{ephemeris}
\end{table*}

\renewcommand\arraystretch{1.2}
\begin{table*}
\caption{\centering The O-C fitting coefficients and mass transfer rate.}
\centering
\small
\begin{tabular}{ccccccccc}
\hline
Parameter& $\Delta{T_0}$& Error& $\Delta{P_0}$& Error& $\beta$&Error & $dM_1/dt$&Error\\ 
&$(\times {10^{-4}} d)$&& $(\times {10^{-7}} d)$&& $(\times {10^{-7}} d$ $ {yr^{-1}})$& & $(\times {10^{-7}} M_\odot$ $ {yr^{-1}})$&\\ 
\hline
BU Tri
&  3.92&25.56& -45.18&10.22& -21.16&2.07&7.26 &0.71\\ 
J1157
&  21.28&61.64& 43.98&26.47& 6.00&5.29&0.62 &0.54\\
 J1236
& 9.11& 29.15& 66.16& 10.01& 8.03& 1.79&0.99 &0.22\\
 J1648
& 1.36& 19.93& -37.04& 4.60& -1.97& 0.46&1.16 &0.27\\ 
J2333
&  -23.88&24.74& -19.76&9.46&  -3.14& 1.79&-2.55 &1.45\\ 
V1232 Her
&  7.75&11.86& -2.44&1.62& -&-&- &-\\
 V1487 Her
& -1.36& 3.11& -14.38& 2.19& -2.89& 0.81& -4.08 & 1.14\\
Z1616& 0.00& 154.22& 11.77& 21.56&- &- &- &-\\
\hline
\end{tabular}
\label{OCtab}
\end{table*}

\vspace{1cm}
\section{Light Curve Solutions}
\label{sec5}
To initiate the light curve solution process, we converted the time data to phase using the new ephemeris provided in Table \ref{ephemeris}. We analyzed the light curves of the target binary systems using version 2.4.9 of the PHysics Of Eclipsing BinariEs (PHOEBE) Python code (\citealt{2016ApJS..227...29P, 2020ApJS..250...34C}). The contact mode was selected for light curve modeling based on the shapes of the observed light curves, the classifications reported in the catalogs, and the systems’ short orbital periods. The gravity-darkening coefficients were set to $g_1=g_2=0.32$ (\citealt{1967ZA.....65...89L}), and the bolometric albedos to $A_1=A_2=0.5$ (\citealt{1969AcA....19..245R}). We adopted the stellar atmosphere model from \cite{2004AA...419..725C}, while the limb-darkening coefficients were left as free parameters during the modeling in PHOEBE.

We subsequently determined initial values for some main parameters to guide the light curve modeling process. The initial estimate of the effective temperature ($T$) was taken from the Gaia DR3 database. This temperature was assigned to the hotter component of the systems based on the depth of the minima observed in the light curves. The initial effective temperature of the cooler star was derived from the observed depth difference between the primary and secondary minima in the light curves.

\begin{figure*}
\centering
\includegraphics[width=0.32\textwidth]{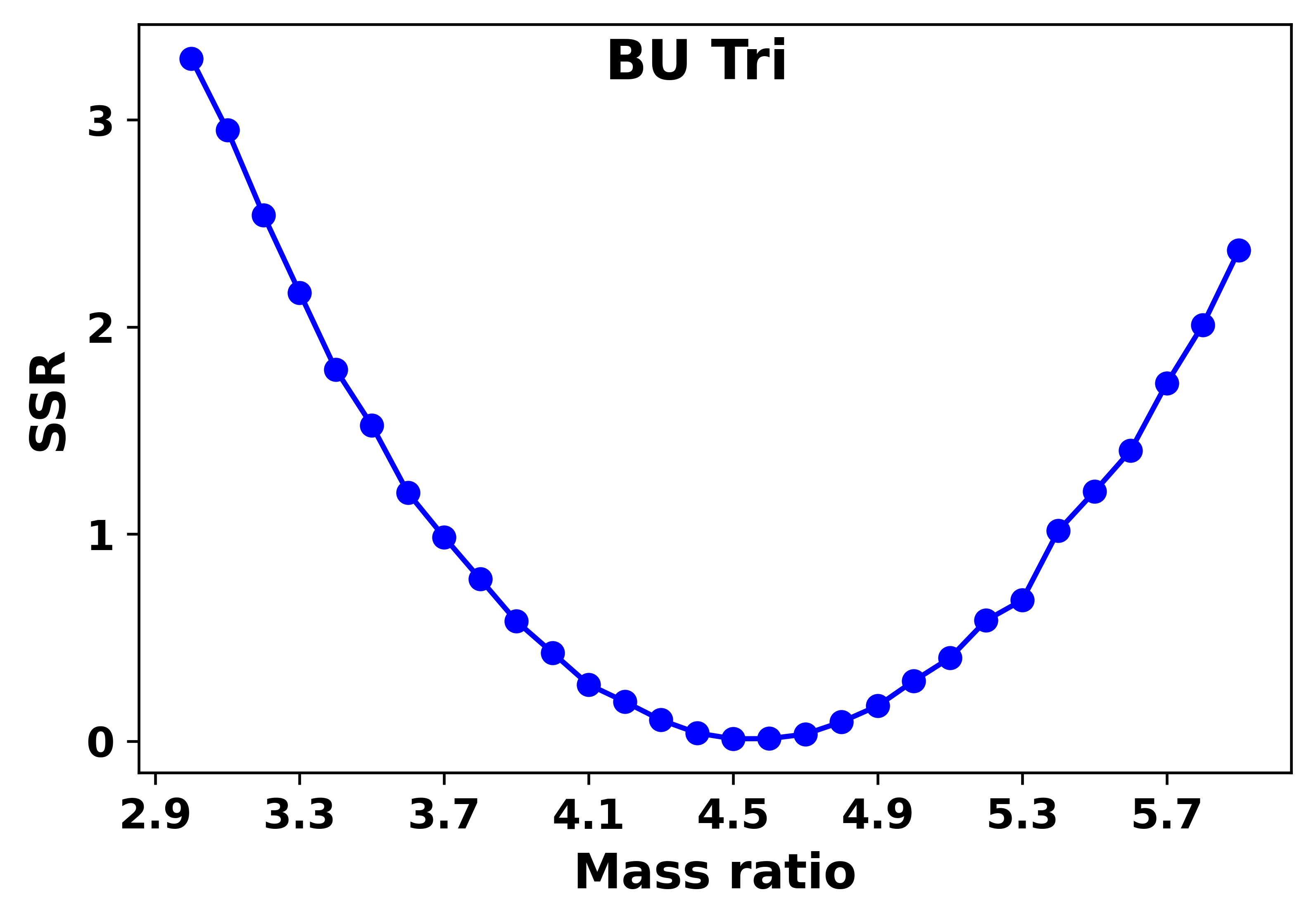}
\includegraphics[width=0.32\textwidth]{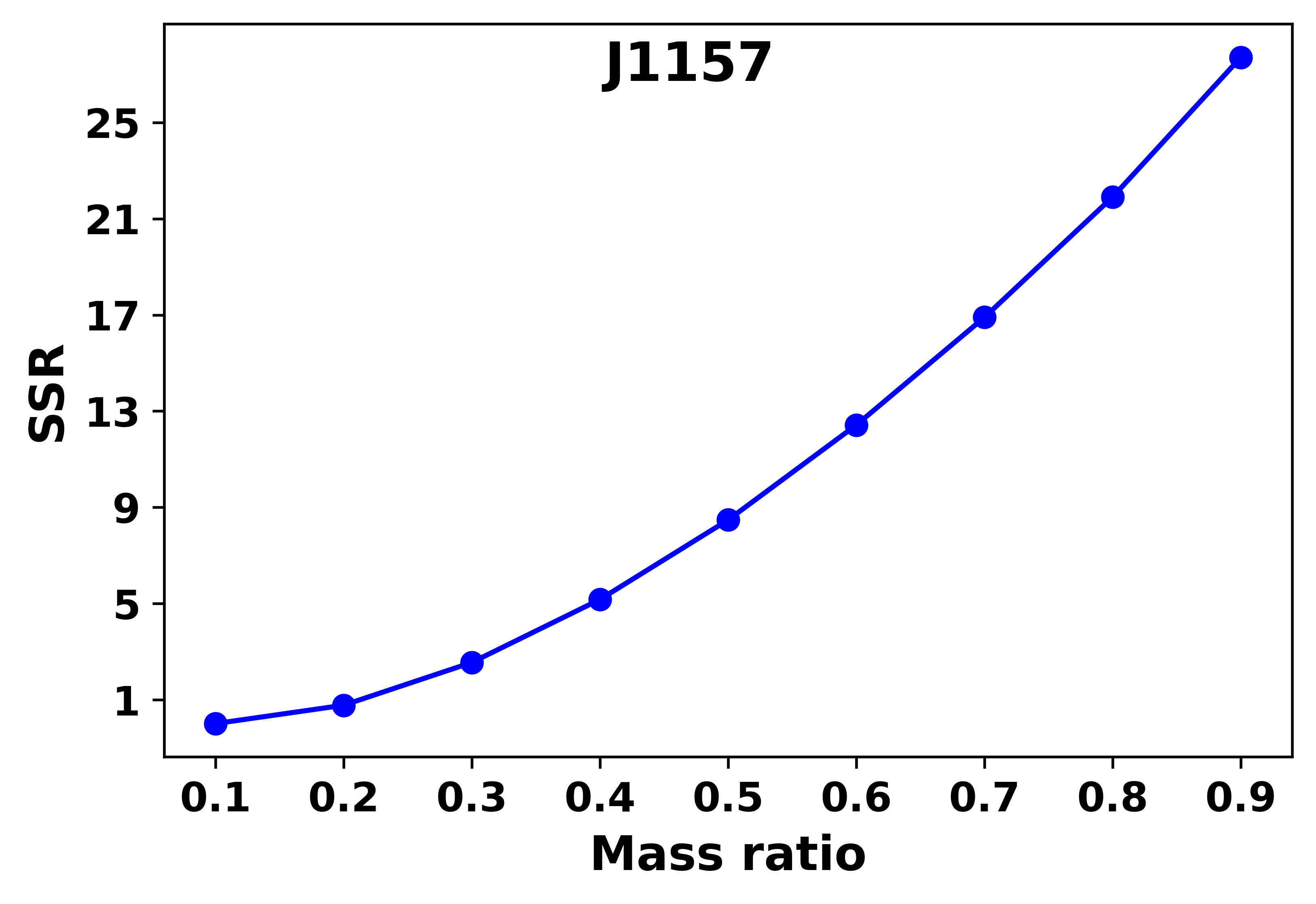}
\includegraphics[width=0.32\textwidth]{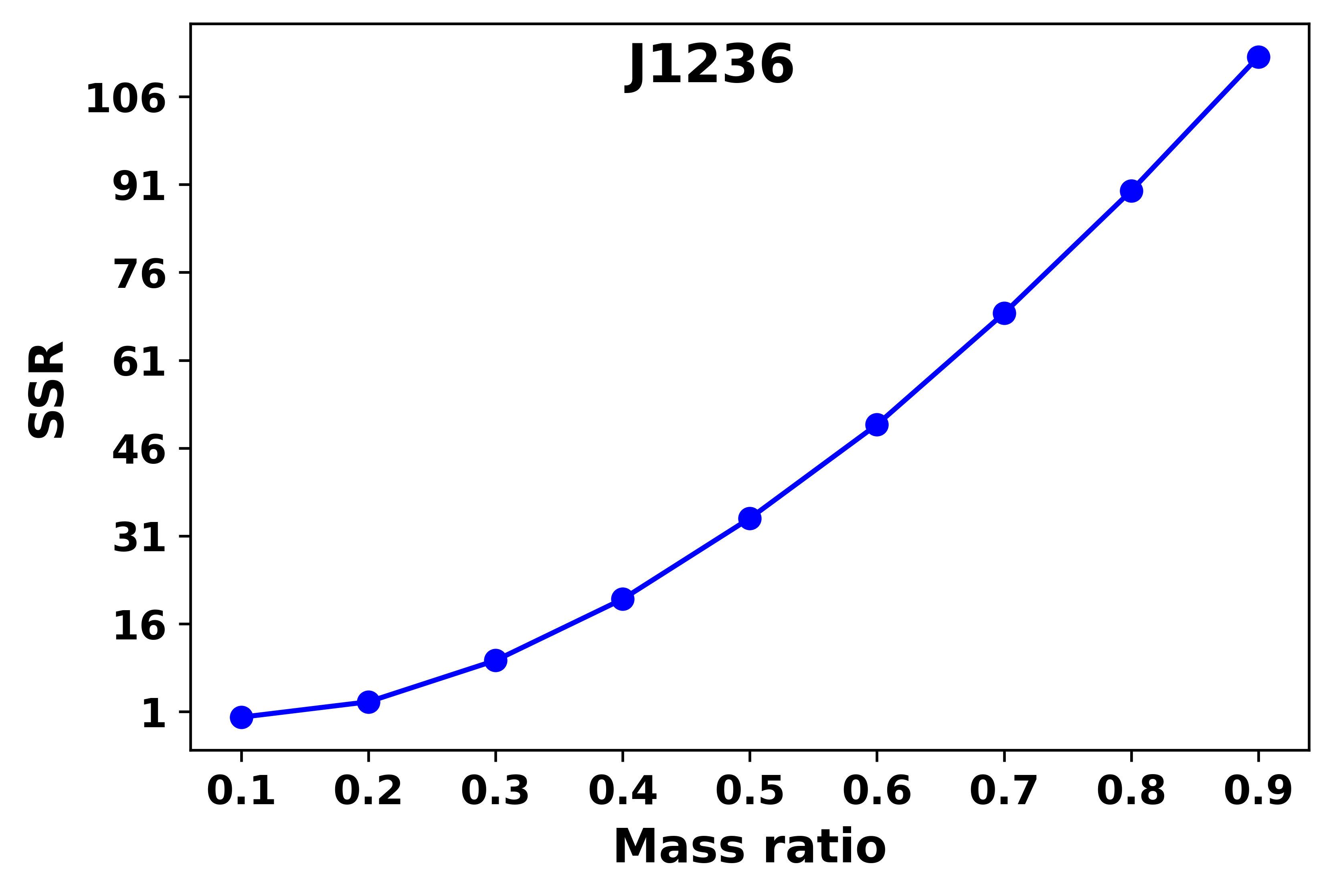}
\includegraphics[width=0.32\textwidth]{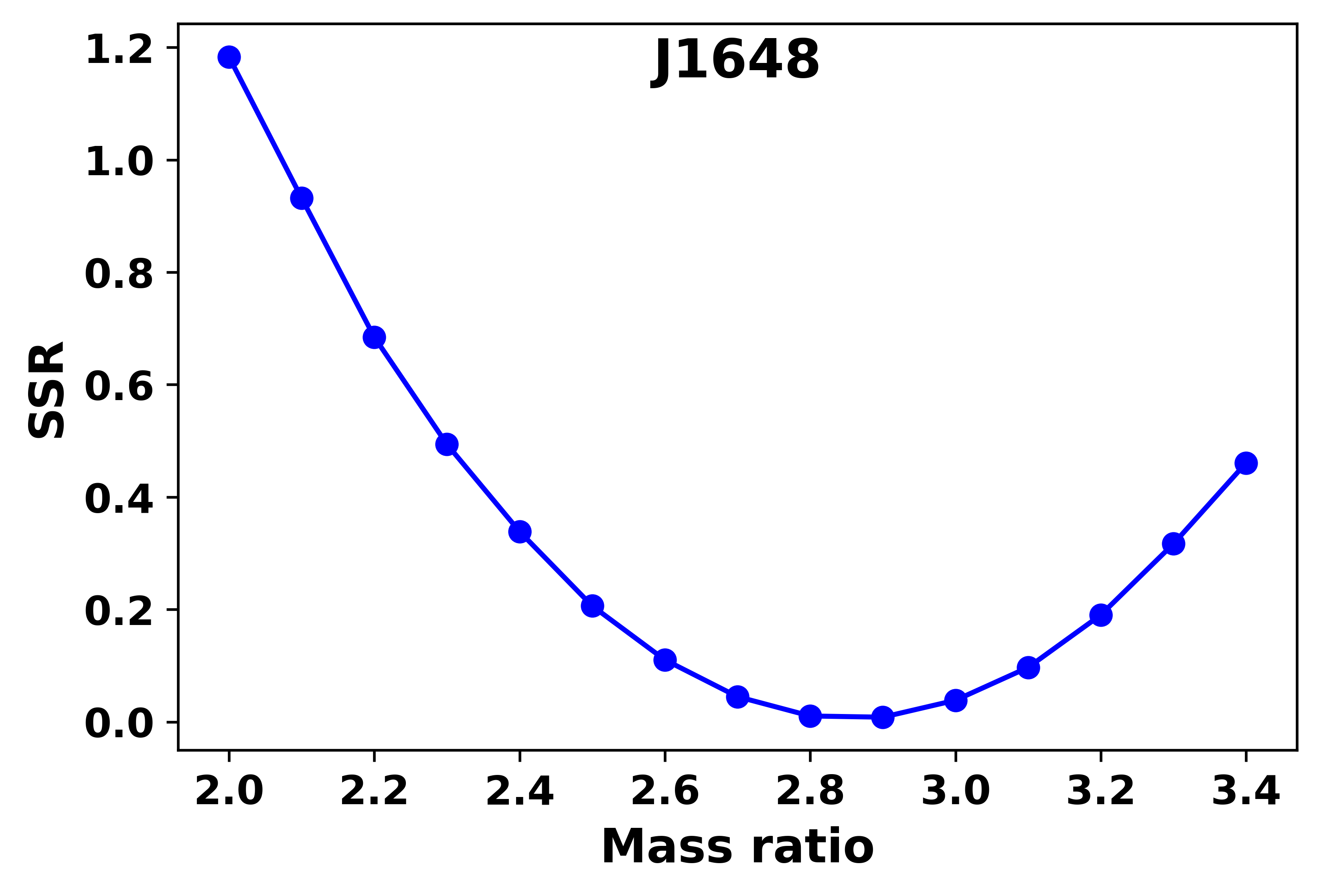}
\includegraphics[width=0.32\textwidth]{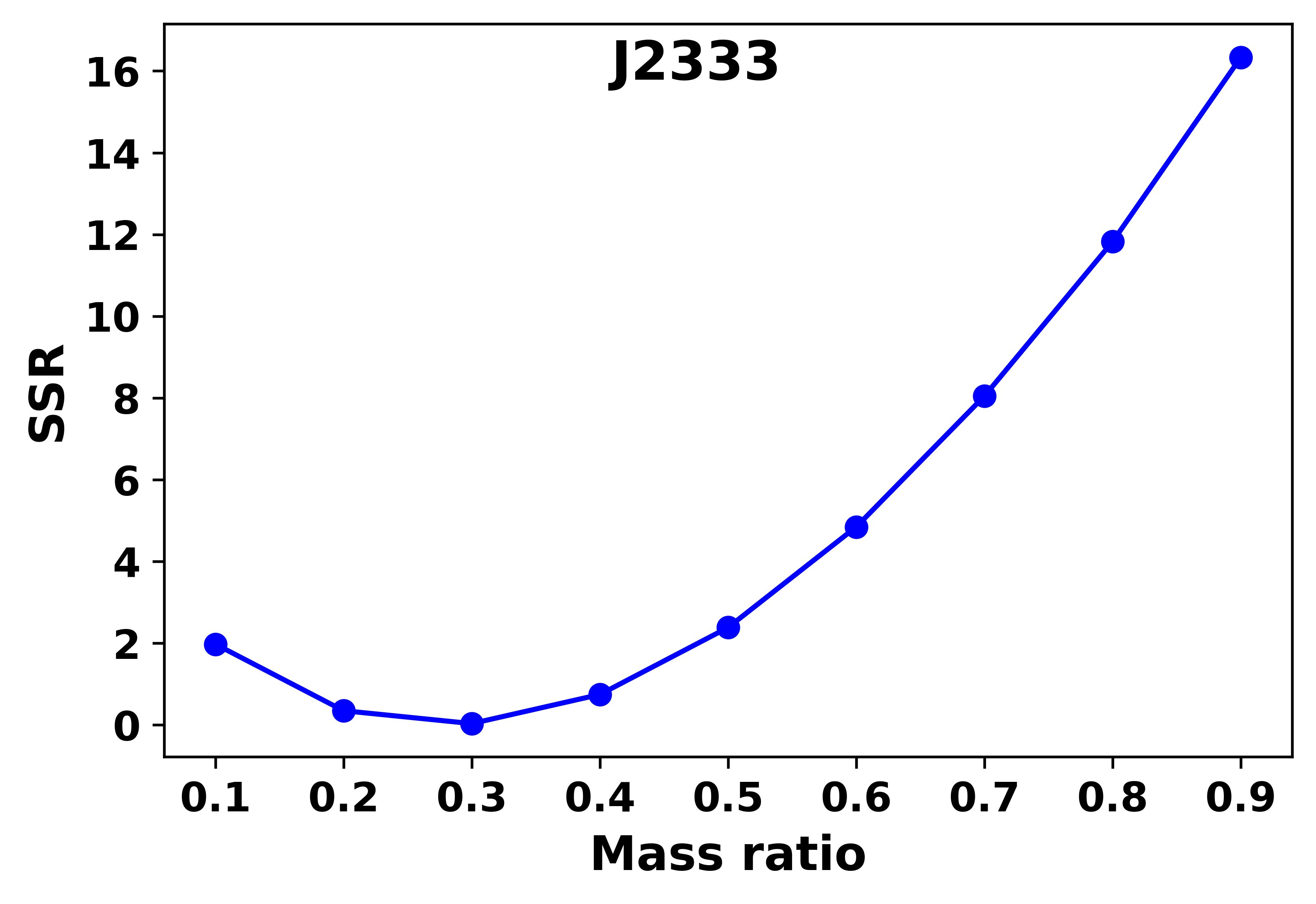}
\includegraphics[width=0.32\textwidth]{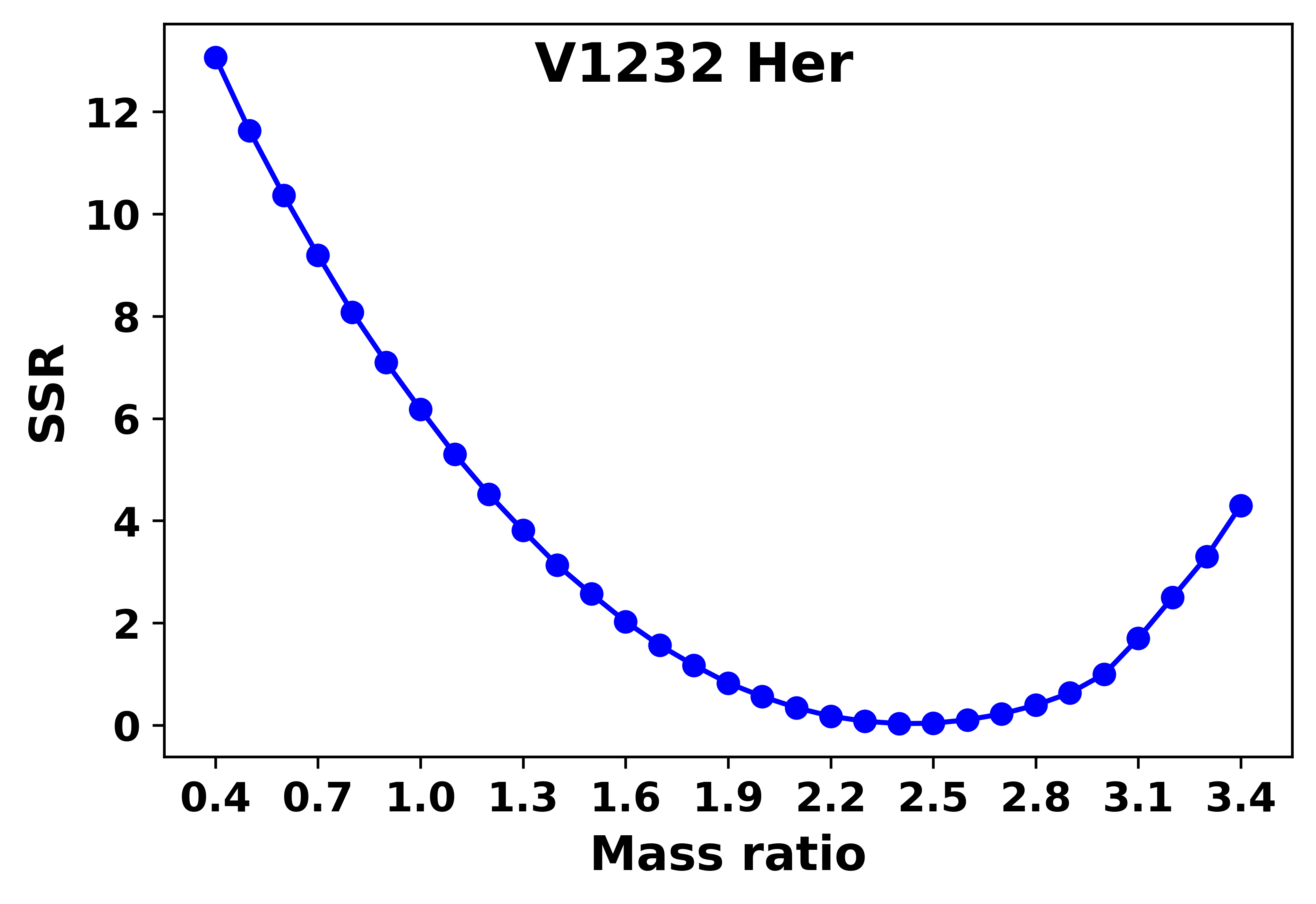}
\includegraphics[width=0.32\textwidth]{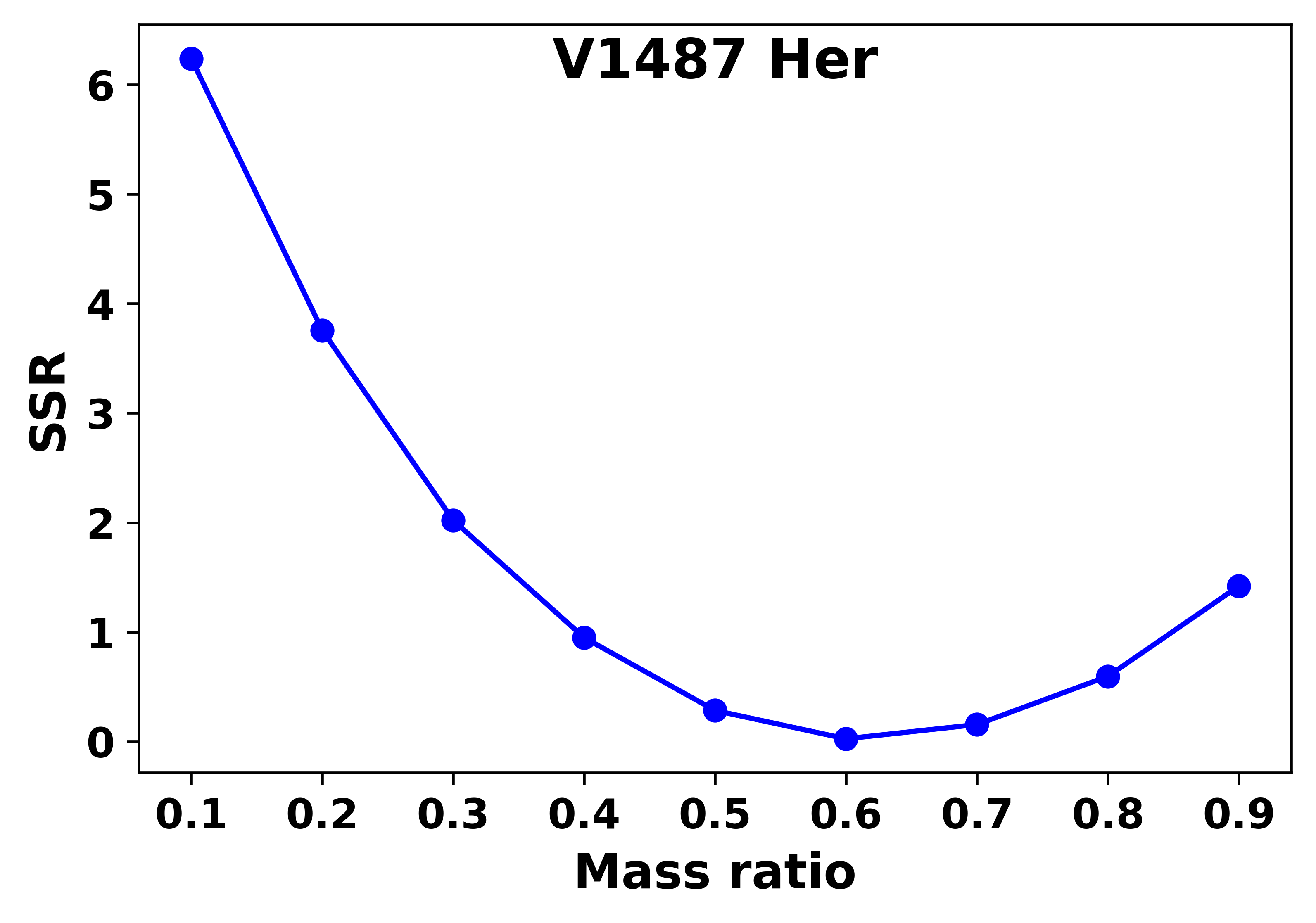}
\includegraphics[width=0.32\textwidth]{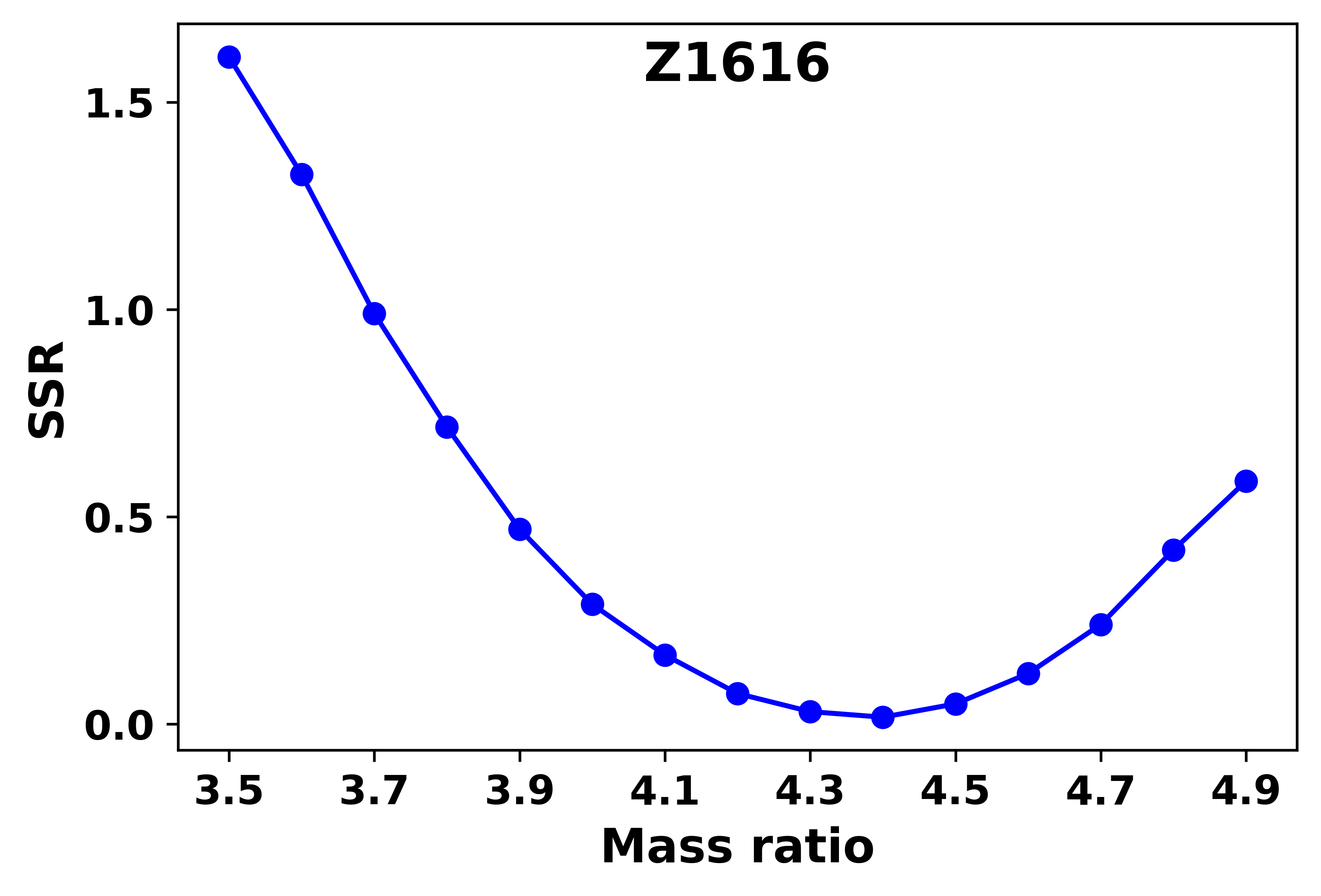}
\caption{The sum of squared residuals as a function of mass ratio.}
\label{q-diagrams}
\end{figure*}

The initial mass ratio ($q$) of the systems was determined using the $q$-search method (\citealt{2005ApSS.296..221T}). The mass ratio range of 0.05 to 12 was explored for target systems. A narrower range was then explored to refine the estimate by minimizing the sum of squared residuals between the observed and synthetic light curves. Figure \ref{q-diagrams} illustrates that each $q$-search curve exhibits a clear minimum sum of squared residuals. Studies such as \cite{2024AJ....168..272P} indicate that estimating the mass ratio parameter using the $q$-search method is more reliable for fully eclipse systems than for partial eclipse binary stars.

Asymmetry in the light curve maxima is a notable feature of many contact binary systems. Six of the target systems show asymmetry in the maxima of their light curves, requiring a cold starspot on one of the components to explain this feature (Table \ref{lc-analysis}). This phenomenon is most plausibly explained by the magnetic activity of the stars, which leads to the formation of starspots, and is referred to as the O'Connell effect (\citealt{1951PRCO....2...85O}, \citealt{2017AJ....153..231S}). While this interpretation is widely used, other physical approaches have also been proposed to explain the phenomenon more comprehensively, including those by \cite{1990ApJ...355..271Z} and \cite{2003ChJAA...3..142L}.

We used the photometric multiband data ($BVR_cI_c$) and initial parameter values to obtain a satisfactory theoretical fit. The optimization tool in PHOEBE was further used to enhance the light curve solution, providing more accurate estimates for the effective temperatures, mass ratio, fillout factor, and orbital inclination. The analysis revealed no evidence of a third light component ($l_3$) in any of the target systems.

The modeling and optimization routines available in PHOEBE do not inherently provide estimates of parameter uncertainties; therefore, we employed the BSN application version 1.0 (\citealt{paki2025bsn}). Designed for Windows operating systems and accessible to members of the BSN project, this application offers substantially higher computational performance in the Markov chain Monte Carlo (MCMC) fitting procedure, generating synthetic light curves more than 40 times faster than PHOEBE. This improvement arises primarily from the application’s optimized architecture and the integration of modern computational libraries, while the core methodologies for analyzing light curves remain consistent with those used in other established binary star modeling packages. For the MCMC simulations, we used BSN with 24 walkers and 1,000 iterations to sample five key parameters ($T_{1,2}$, $q$, $f$, and $i$), from which uncertainty estimates were obtained, and the average upper and lower bounds of these uncertainties are presented in Table \ref{lc-analysis}. Notably, the final parameter estimates and synthetic light curves produced by BSN and PHOEBE for the systems studied were effectively equivalent.

\begin{figure*}
\centering
\includegraphics[width=0.86\textwidth]{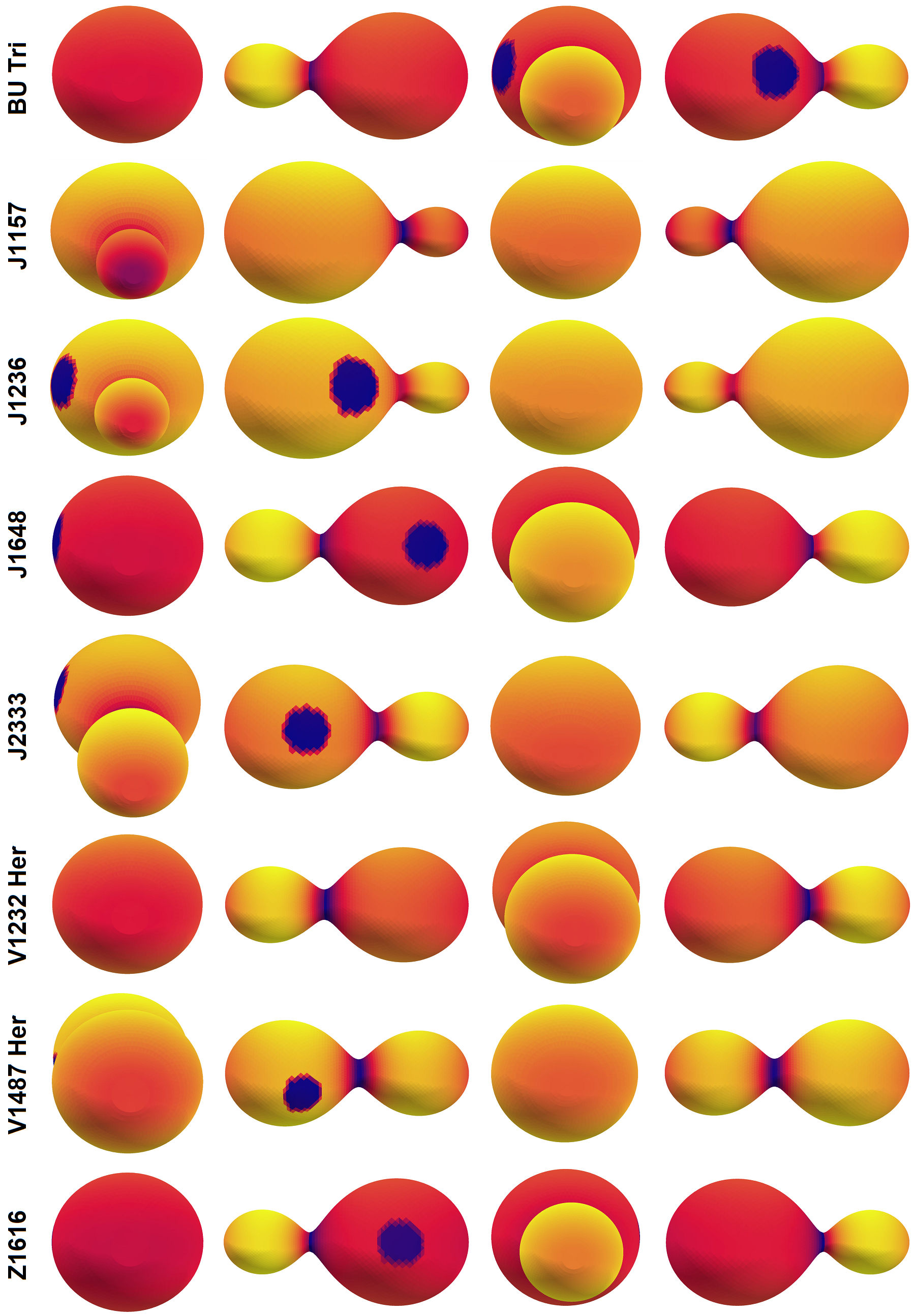}
\caption{Three-dimensional views of the stars in the target binary systems at orbital phases 0.0, 0.25, 0.5, and 0.75, respectively.}
\label{3d}
\end{figure*}

The final results of the light curve analysis are present in Table \ref{lc-analysis}, including the starspot parameters: colatitude (Col.$^\circ$), longitude (Long.$^\circ$), angular radius (Radius$^\circ$), and the temperature ratio ($T_{spot}/T_{star}$). The three-dimensional representations of the binary systems, based on the final model parameters, are shown in Figure \ref{3d}.

\begin{table*}
\renewcommand\arraystretch{1.8}
\caption{Light curve solutions of the target binary stars.}
\centering
\begin{center}
\footnotesize
\begin{tabular}{c c c c c c c c c}
\hline
Parameter & BU Tri & J1157 & J1236 & J1648 & J2333 & V1232 Her & V1487 Her & Z1616\\
\hline
$T_{1}$ (K) & 5846(55) & 6969(45) &	5658(23) & 5798(37) & 4990(27) & 5334(19) & 4901(34) & 5323(22)\\
$T_{2}$ (K) & 5509(60) & 6697(40) &	5598(17) & 5337(31) & 5057(27) & 5197(21) & 4879(30) & 4911(20)\\
$q=M_2/M_1$ & 4.527(43) & 0.095(5) & 0.099(1) & 2.860(44) & 0.277(17) & 2.443(42) & 0.619(35) & 4.378(24)\\
$i^{\circ}$ & 83.46(51) & 78.96(43) & 81.11(56) & 82.32(31) & 71.50(21) &	81.90(44) & 83.02(25) & 85.62(34)\\
$f$ & 0.371(41) & 0.386(51) & 0.572(40) & 0.178(16) & 0.309(17) & 0.291(14) & 0.195(9) & 0.163(13)\\
$\Omega_1=\Omega_2$ & 8.340(236) & 1.920(38) & 1.919(28) & 6.321(109) & 2.361(54) & 5.690(177) & 3.028(114) & 8.286(103)\\
$l_1/l_{tot}$($V$) & 0.262(3) &	0.901(19) &	0.883(23) & 0.370(3) & 0.740(14) & 0.344(3) & 0.612(4) & 0.297(3)\\
$l_2/l_{tot}$($V$) & 0.738(2) &	0.099(5) & 0.117(5) & 0.630(6) & 0.260(5) & 0.656(4) & 0.388(5) & 0.703(5)\\
$r_{(mean)1}$ &	0.278(20) &	0.595(11) &	0.597(13) & 0.304(12) & 0.511(16) & 0.325(21) & 0.438(16) & 0.269(9)\\
$r_{(mean)2}$ &	0.532(17) &	0.216(14) &	0.225(17) & 0.485(11) & 0.293(18) & 0.479(19) & 0.355(16) & 0.519(7)\\
\hline
$Col.^\circ$(spot) & 99 & - & 81 & 98 & 75 & - & 104 & 94\\
$Long.^\circ$(spot) & 301 &	- &	303 & 106 &	281 & - & 290 & 89\\
$Radius^\circ$(spot) & 19 &	- &	18 & 18 & 17 & - &	18 & 16\\
$T_{spot}/T_{star}$ & 0.88 & - & 0.83 & 0.87 & 0.88 & - & 0.89 & 0.90\\
Component &	Secondary &	- &	Primary & Secondary & Primary & - & Primary & Secondary\\
\hline
\end{tabular}
\end{center}
\label{lc-analysis}
\end{table*}

\begin{figure*}
\centering
\includegraphics[width=0.495\textwidth]{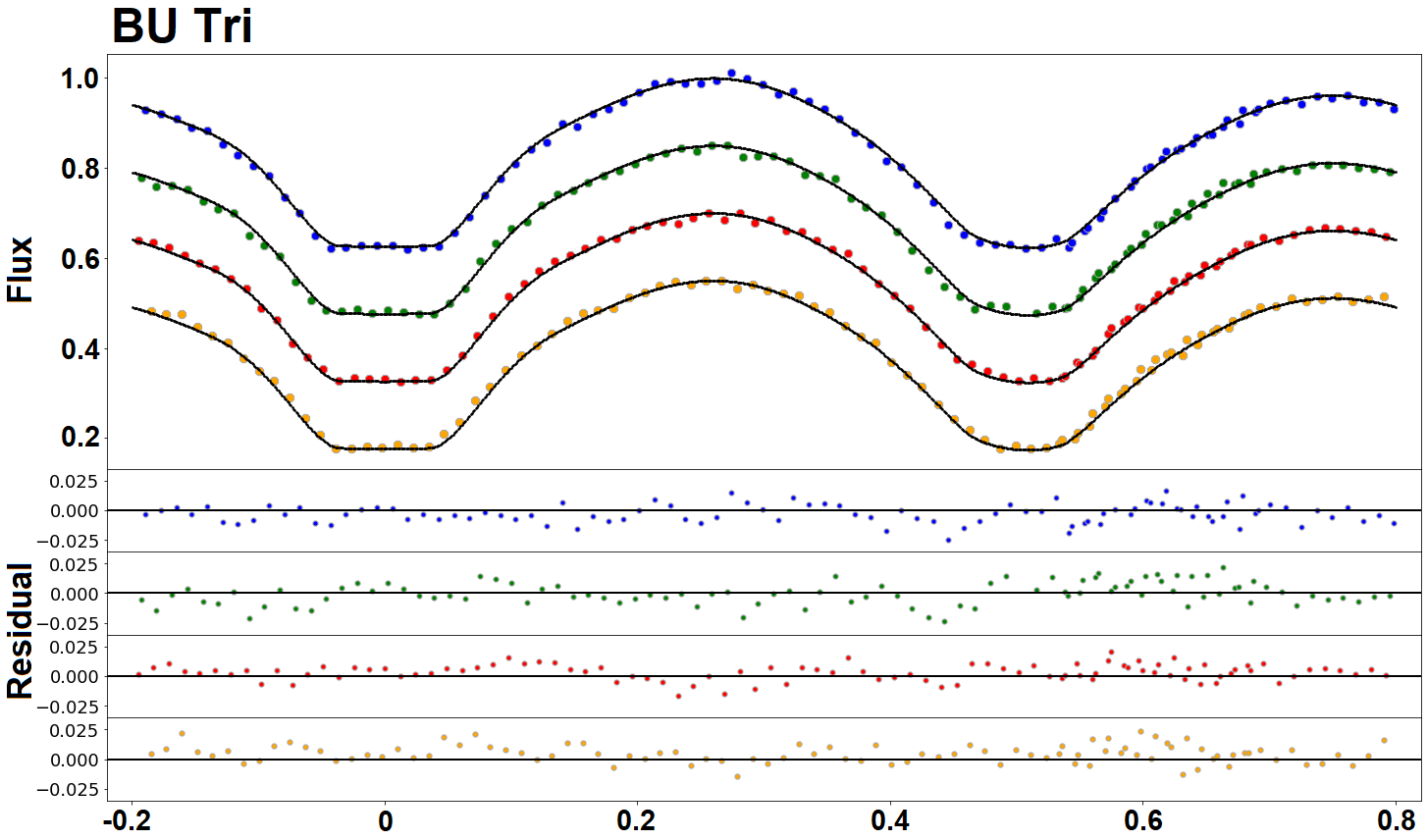}
\includegraphics[width=0.495\textwidth]{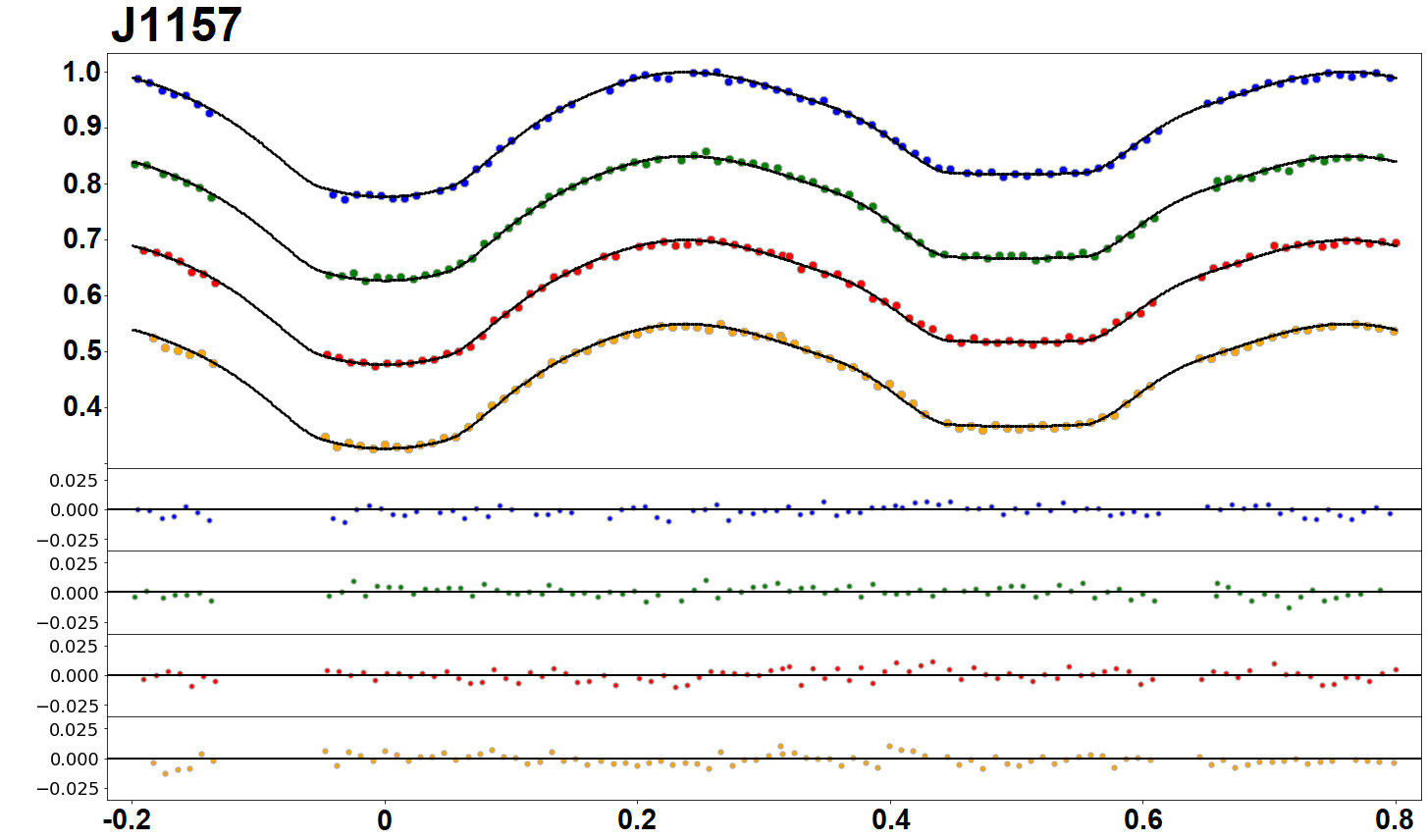}
\includegraphics[width=0.495\textwidth]{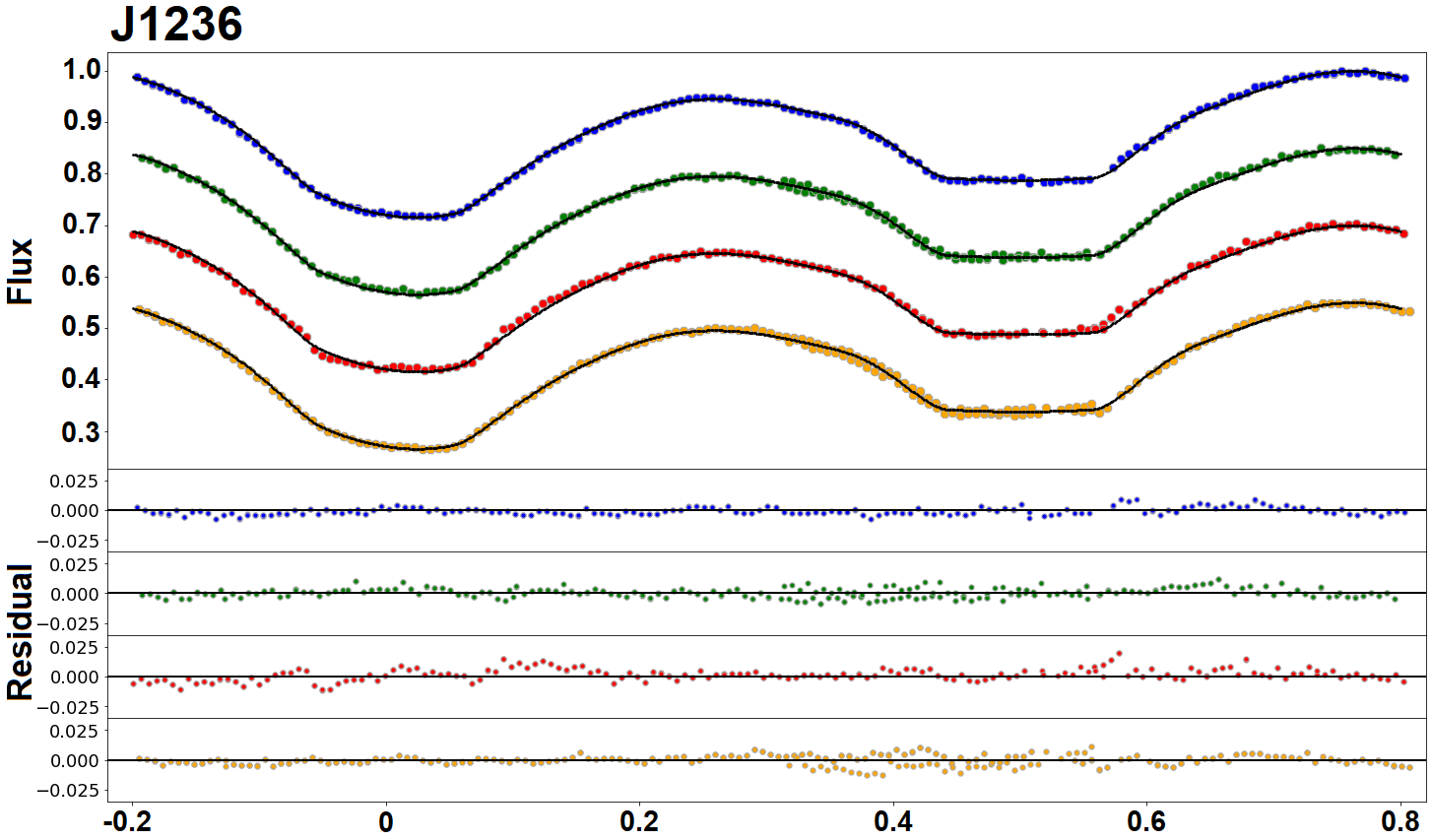}
\includegraphics[width=0.495\textwidth]{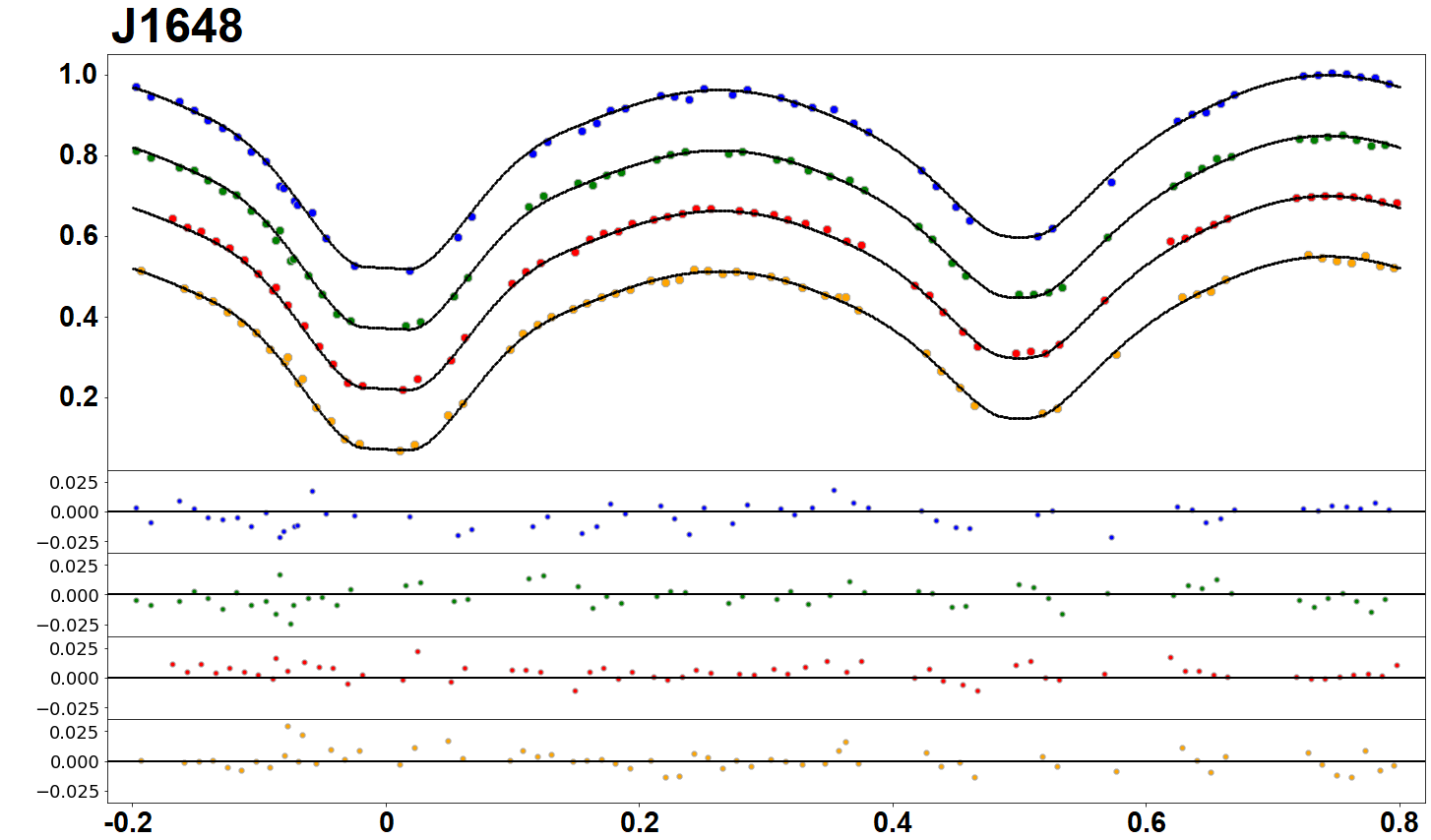}
\includegraphics[width=0.495\textwidth]{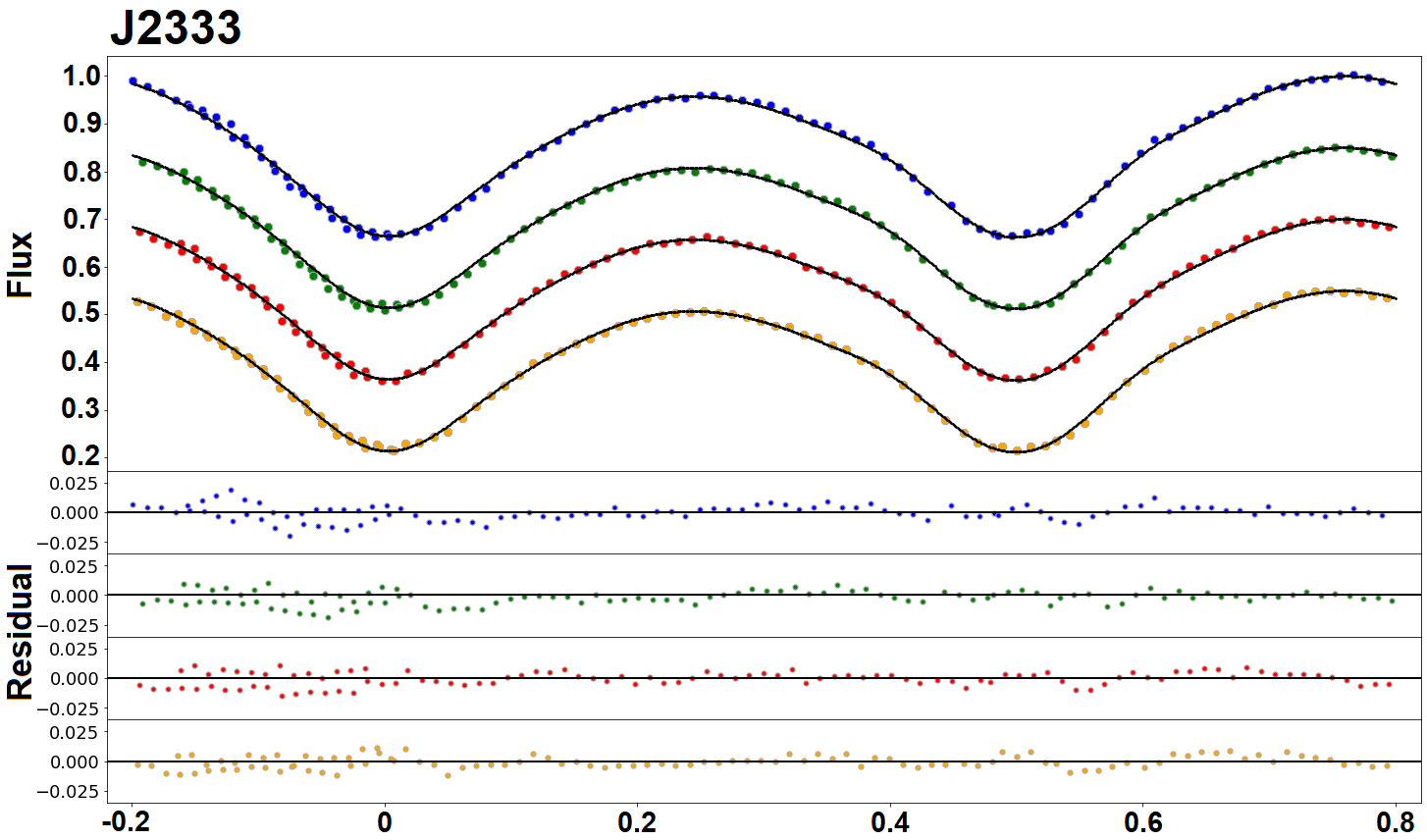}
\includegraphics[width=0.495\textwidth]{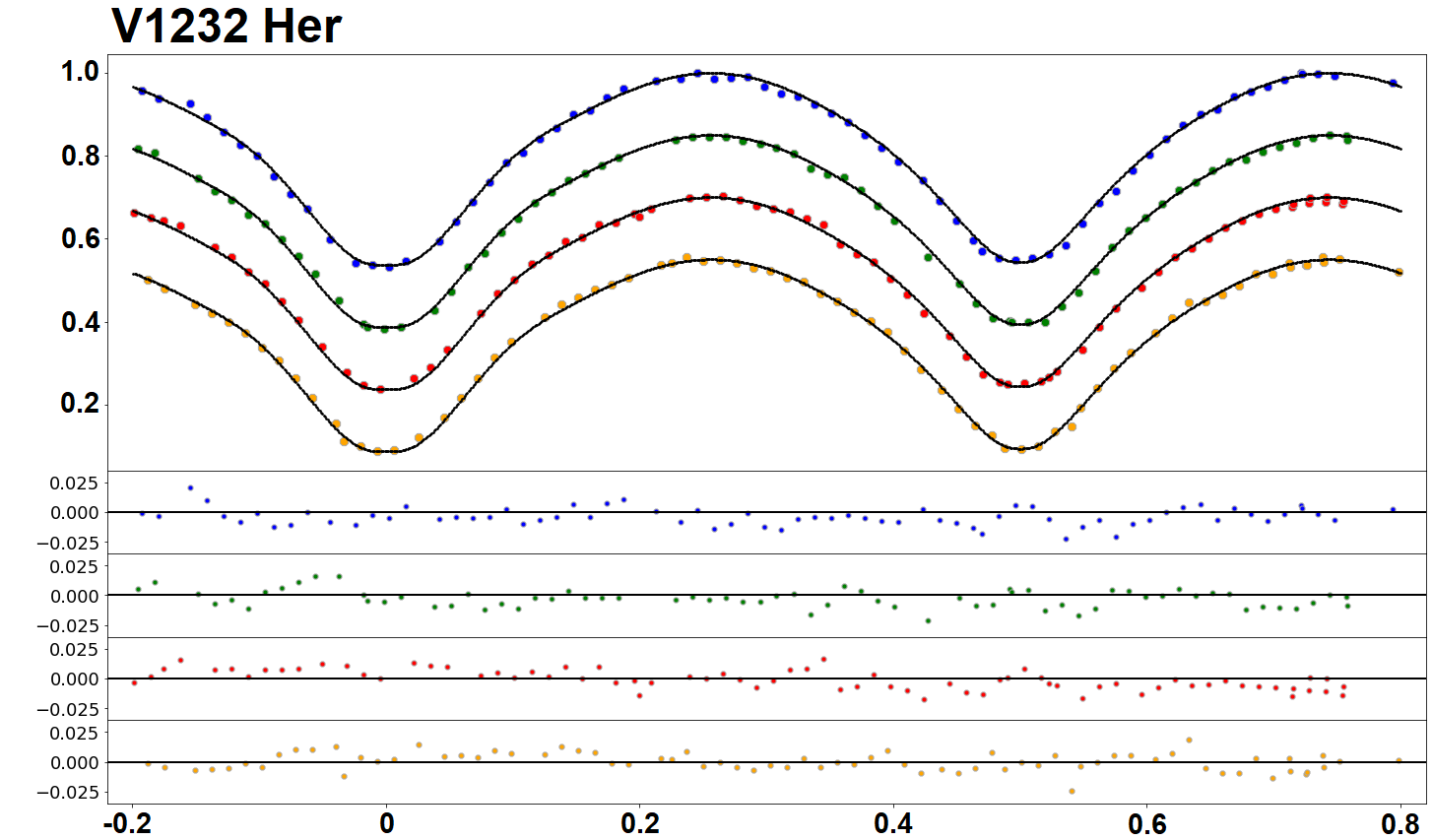}
\includegraphics[width=0.495\textwidth]{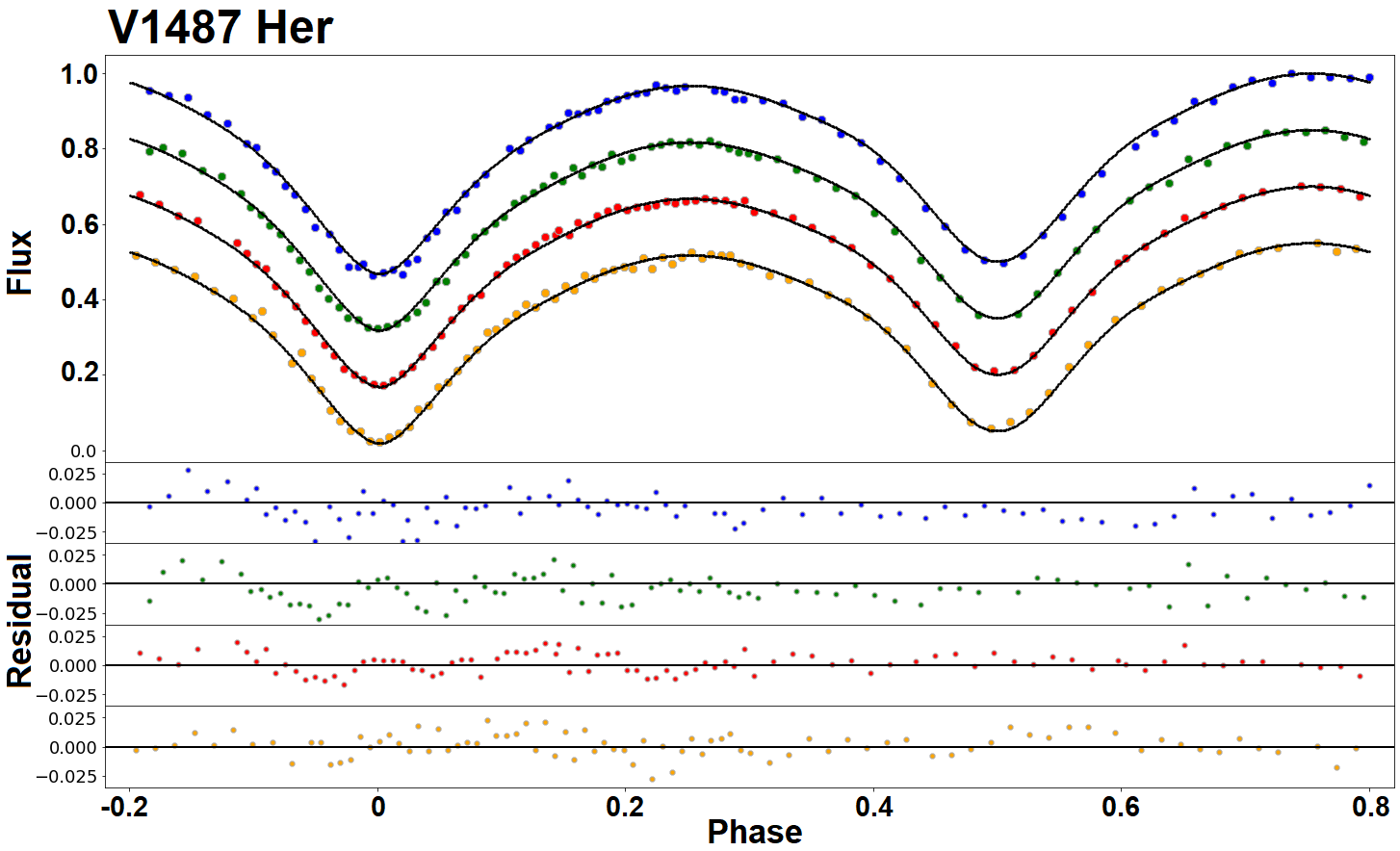}
\includegraphics[width=0.495\textwidth]{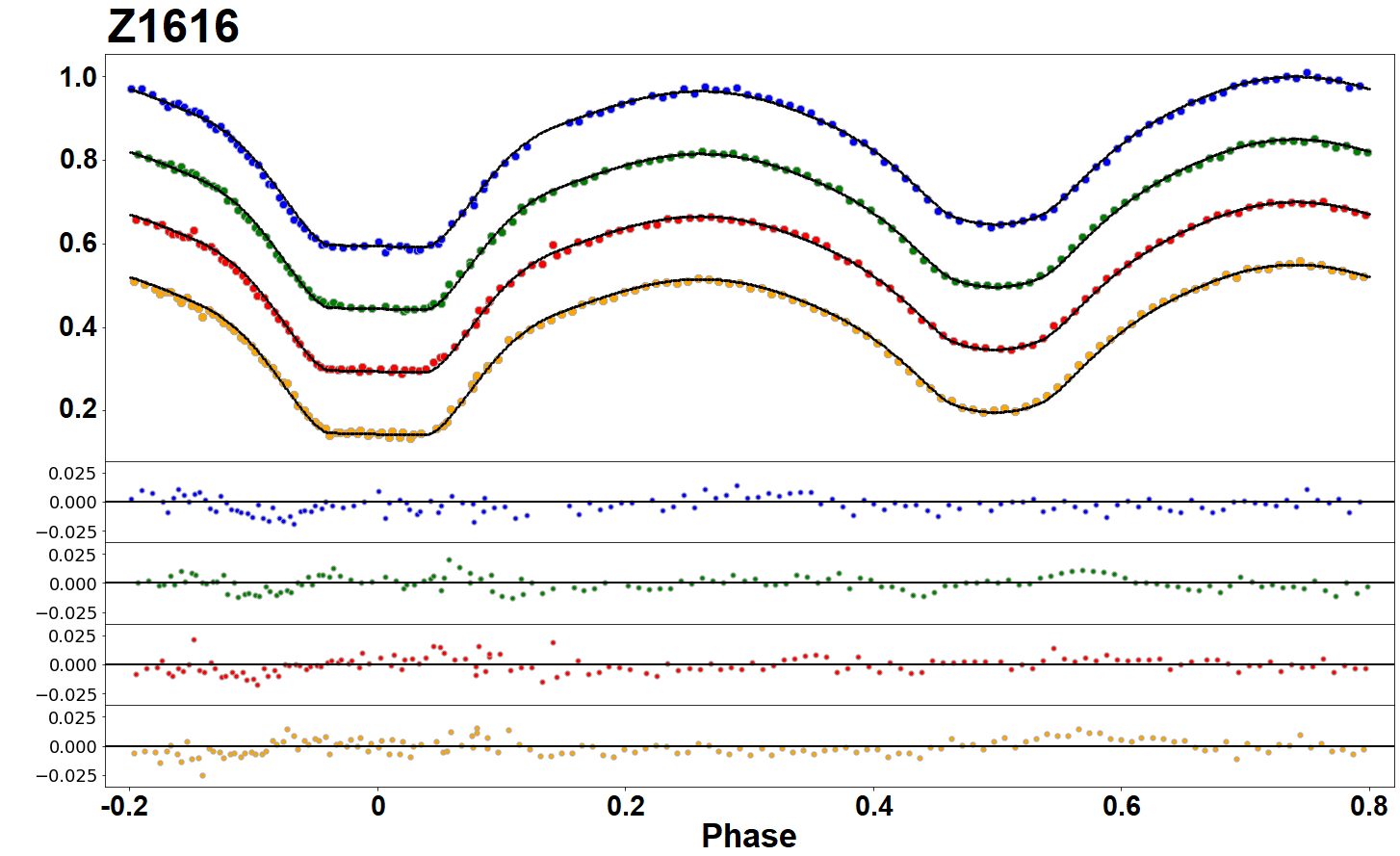}
\caption{The colored dots represent the observed light curves of the systems in different filters, and the synthetic light curves, generated using the light curve solutions, are also shown. Residuals are shown at the bottom of each panel}.
\label{LCs}
\end{figure*}

\vspace{0.6cm}
\section{Fundamental Parameters}
\label{sec6}
We used the Gaia DR3 parallax to estimate the absolute parameters of the target systems, making this method a reliable option when only photometric data is available (\citealt{2024NewA..11002227P}). To ensure suitability of the Gaia DR3 parallax method for our target systems, we first calculated and reviewed the interstellar extinction ($A_V$) parameter. According to the \citealt{2024PASP..136b4201P} study, the $A_V$ value should be lower than approximately 0.4. Using the 3D dust map from \citet{2019ApJ...887...93G}, we calculated the $A_V$ values and confirmed that the target systems fall within the acceptable range (Table \ref{absolute}).

This method estimates the system's absolute magnitude ($M_V$) from the maximum brightness of the system $V_{max}$, Gaia DR3 distance, and $A_V$. We utilized the $V_{max}$ values obtained from the observational data presented in this study (Table \ref{absolute}). Subsequently, the $l_{1,2}/l_{tot}$ parameter derived from the $V$ filter in the light curve solution process was used to calculate $M_{V1}$ and $M_{V2}$. The absolute bolometric magnitudes ($M_{bol_1,2}$) were estimated using bolometric corrections ($BC_{1,2}$) derived from \cite{1996ApJ...469..355F}. Then, the stellar radii in the binary systems were estimated using the relationship between absolute bolometric magnitude ($M_{bol}$) and luminosity ($L$). We adopted $M_{bol\odot}=4.73$ mag., as reported by \cite{2010AJ....140.1158T}, throughout the estimation process. Moreover, using the luminosity and the effective temperatures from the light curve solution, the stellar radii ($R$) can be calculated. It is important to note that this radius corresponds to the volume-equivalent radius of the surface defined by the potential shell. The semi-major axis $a$ of each system is determined using $r_{mean1,2}$ from the light curve solution and $R_{1,2}$, followed by averaging $a_1$ and $a_2$. By utilizing the parameters $a$, $P$, and $q$, the masses of the individual components can be computed through the application of Kepler's third law:

\begin{eqnarray}
M{_1}=\frac{4\pi^2a^3}{GP^2(1+q)}\label{eq:M1},\\
M{_2}=q\times{M{_1}}\label{eq:M2}.
\end{eqnarray}

Furthermore, the orbital angular momentum ($J_0$) for each system was calculated using the total mass of the system, $q$, and $P$ (\citealt{2006MNRAS.373.1483E}), and the results are presented in Table \ref{absolute}.

\begin{equation}\label{eqJ0}
J_0=\frac{q}{(1+q)^2} \sqrt[3] {\frac{G^2}{2\pi}M^5P}.
\end{equation}

The estimated absolute parameters for the target binary systems are listed in Table \ref{absolute}.

\begin{table*}
\renewcommand\arraystretch{1.8}
\caption{Estimated absolute parameters of the systems.}
\centering
\begin{center}
\footnotesize
\begin{tabular}{c c c c c c c c c}
\hline
Parameter & BU Tri & J1157 & J1236 & J1648 & J2333 & V1232 Her & V1487 Her & Z1616\\
\hline
$M_1(M_\odot)$ 	&	0.237(46)	&	1.000(154)	&	1.006(32)	&	0.361(34)	&	1.869(107)	&	0.238(3)	&	0.587(62)	&	0.195(19)	\\
$M_2(M_\odot)$ 	&	1.072(209)	&	0.095(14)	&	0.100(3)	&	1.034(97)	&	0.518(29)	&	0.582(8)	&	0.363(39)	&	0.854(81)	\\
$R_1(R_\odot)$ 	&	0.563(62)	&	1.264(117)	&	1.168(76)	&	0.656(41)	&	1.275(84)	&	0.530(29)	&	0.671(53)	&	0.480(25)	\\
$R_2(R_\odot)$ 	&	1.097(132)	&	0.456(42)	&	0.437(24)	&	1.059(65)	&	0.726(47)	&	0.787(45)	&	0.541(39)	&	0.934(53)	\\
$L_1(L_\odot)$ 	&	0.332(62)	&	3.379(543)	&	1.255(142)	&	0.436(43)	&	0.904(99)	&	0.204(20)	&	0.233(31)	&	0.166(15)	\\
$L_2(L_\odot)$ 	&	0.995(198)	&	0.375(60)	&	0.168(16)	&	0.817(82)	&	0.309(33)	&	0.405(41)	&	0.149(18)	&	0.455(45)	\\
$M_{bol1}(mag.)$ 	&	5.936(198)	&	3.418(173)	&	4.493(123)	&	5.642(107)	&	4.849(119)	&	6.468(103)	&	6.323(143)	&	6.693(99)	\\
$M_{bol2}(mag.)$ 	&	4.745(214)	&	5.805(174)	&	6.676(103)	&	4.959(109)	&	6.014(118)	&	5.722(107)	&	6.807(133)	&	5.594(106)	\\
$log(g)_1(cgs)$ 	&	4.312(179)	&	4.235(147)	&	4.306(70)	&	4.362(95)	&	4.499(82)	&	4.366(53)	&	4.553(115)	&	4.366(87)	\\
$log(g)_2(cgs)$ 	&	4.388(188)	&	4.098(146)	&	4.157(62)	&	4.403(95)	&	4.431(80)	&	4.411(56)	&	4.532(110)	&	4.429(91)	\\
$a(R_\odot)$ 	&	2.043(130)	&	2.118(107)	&	1.949(21)	&	2.171(68)	&	2.487(47)	&	1.637(7)	&	1.528(54)	&	1.792(56)	\\
$logJ_0(cgs)$ 	&	51.283(140)	&	50.902(131)	&	50.906(27)	&	51.450(65)	&	51.776(56)	&	51.073(7)	&	51.213(83)	&	51.119(67)	\\
\hline
$V_{max}(mag.)$ 	&	15.61(7)	&	14.62(6)	&	14.32(10)	&	14.43(8)	&	14.72(9)	&	14.68(8)	&	15.67(9)	&	14.73(7)	\\
$A_V(mag.)$ 	&	 0.139(1) 	&	 0.053(1) 	&	0.079(1) 	&	 0.040(1) 	&	 0.298(1) 	&	 0.054(1) 	&	 0.188(1) 	&	 0.138(1)	\\
$BC_1(mag.)$	&	-0.068(9)	&	0.030(1)	&	-0.102(5)	&	-0.076(6)	&	-0.312(13)	&	-0.183(6)	&	-0.355(17)	&	-0.186(7)	\\
$BC_2(mag.)$ 	&	-0.135(14)	&	0.019(2)	&	-0.114(3)	&	-0.182(10)	&	-0.283(11)	&	-0.228(7)	&	-0.366(15)	&	-0.350(9)	\\
\hline
\end{tabular}
\end{center}
\label{absolute}
\end{table*}

\vspace{0.6cm}
\section{Discussion and Conclusion}
\label{sec7}
In this work, we present the photometric analysis, including light curve modeling, orbital period variation study, and determination of absolute parameters for eight contact binary systems. The data were obtained through multiband photometric observations carried out at the San Pedro Mártir Observatory in México. The outcomes of the analysis form the basis for the subsequent discussion and conclusions:

A) A long-term increase in the orbital period is typically driven by mass transfer from the less massive component to the more massive one. Conversely, a decrease in the long-term orbital period is often due to angular momentum loss or mass transfer from the more massive component to the less massive one. The following equation, as described by \cite{k1958}, can be used to calculate the mass transfer rate by assuming fully conservative mass transfer without angular momentum loss,

\begin{equation}\label{7}
\begin{aligned}
\frac{\dot{P}}{P}=-3\dot{M}(\frac{1}{M_1}-\frac{1}{M_2}).
\end{aligned}
\end{equation}

The derived mass transfer rates are displayed in Table \ref{OCtab}. For cases exhibiting a long-term increase in orbital period, the derived mass transfer rates represent lower limits, as potential angular momentum losses through gravitational radiation and/or magnetic braking are not accounted for. Conversely, in systems showing orbital period decrease, the calculated mass transfer rates should be considered upper limits, as the period is further shrunk by angular momentum loss processes.

Z1616 shows cyclic variation in the orbital period. Such variation can result from the magnetic activity of one or both components \citep{1992ApJ...385..621A} or the light travel time effect (LTTE) due to an additional component \citep{2016ApJ...817..133Z}. If the cyclic variation results from magnetic activity, we calculated the variation in the magnetic quadruple moment ($\Delta Q$) using the equation from \cite{1992ApJ...385..621A},

\begin{equation}
\frac{\Delta P}{P} = \frac{2\pi \times A}{P_{mod} } = -9  (\frac{R}{a} )^{2}  \frac{\Delta Q}{M R^{2} },
\end{equation}

\noindent where $M$ and $R$ represent the mass and the radius of the active component, while $a$ means the semi-major axis of the binary. $\Delta Q_{1}=1.90 \times 10^{49}\, g \,\text{cm}^{2}$ and $\Delta Q_{2}=8.33 \times 10^{49}\, g \,\text{cm}^{2}$ were calculated for the two components. The derived values are substantially lower than the typical values of  $10^{51} \sim  10^{52}\, g \,\text{cm}^{2}$ in close binaries \citep{1999A&A...349..887L}. This significant discrepancy demonstrates that the Applegate mechanism cannot adequately account for the observed periodic variation in this system. Therefore, the cyclic modulation can be caused by LTTE via the third companion. The mass function of the tertiary companion, $f(M_{3})$, was determined the orbital dynamics relationship expressed as

\begin{equation}
f(M_{3}) = \frac{(M_{3}\sin i_{3})^{3} }{(M_{1}+M_{2}+M_{3})^{2} } = \frac{4\pi^{2}}{GP_{3}^{2}}\times (a_{12}\sin i_{3})^3,
\end{equation}

\noindent we determined that $f(M_{3})=0.031(\pm0.083)$ $M_{\odot}$ and $a_{12}\sin i_{3}=1.50\pm1.36$ AU. If the orbital inclination of the tertiary component is $i=90^\circ$, the minimum mass of the tertiary component, $M_{3min}$ = 0.040($\pm0.075$) $M_{\odot}$, and the maximum distance to the mass center of the triple system, $a_{3max}=4.66\pm9.67$ AU were derived. We think the tertiary component is a brown dwarf. Because the time span of the O-C curve of this system is not long enough, more observations are needed in the future to confirm this result.

B) In six of the analyzed contact binary systems, asymmetries between the light curve maxima necessitated the inclusion of a cool starspot on one of the stellar components. These asymmetries are characteristic of the O'Connell effect, that affects the shape and symmetry of light curves in contact binaries (\citealt{1951PRCO....2...85O}). The effective temperatures of the target stellar components range from 4879 K to 6969 K. The component temperature difference was smallest in V1487 Her at 22 K, while J1648 exhibited the largest, reaching 461 K. Table \ref{conclusion} lists the temperature differences ($\Delta T = T1 - T2$) for each system. Spectral categories were determined based on the temperature criteria provided by \cite{2000asqu.book.....C} and \cite{2018MNRAS.479.5491E} (Table \ref{conclusion}).

\begin{table*}
\renewcommand\arraystretch{1.5}
\caption{Some conclusions regarding the target systems.}
\centering
\begin{center}
\footnotesize
\begin{tabular}{c c c c c c c c c}
\hline
Parameter & BU Tri & J1157 & J1236 & J1648 & J2333 & V1232 Her & V1487 Her & Z1616\\
\hline
$\Delta T=T_1-T_2$ ($K$) 	&	337	&	272	&	60	&	461	&	-67	&	137	&	22	&	412	\\
Spectral category & G3-G8 & F1-F3 & G7-G8 & G3-K0 & K1-K1 & K0-K0 & K2-K2 & K0-K2 \\
$\Delta a=a_1-a_2$ ($R_\odot$)	&	0.037	&	0.013	&	0.014	&	0.026	&	0.017	&	0.012	&	0.008	&	0.016	\\
$f$ category & medium & medium & deep & shallow & medium & medium & shallow & shallow\\
Subtype & W & A & A & W & W & W & A & W \\
$M_{1i}$ ($M_{\odot}$) & 0.610 & 0.423 & 0.520 & 0.632 & 1.676 & 0.180 & 0.342 & 0.441\\
$M_{2i}$ ($M_{\odot}$) & 1.613 & 1.812 & 1.548 & 1.557 & 1.093 & 1.434 & 1.093 & 1.423\\
$M_{lost}$ ($M_{\odot}$) & 0.914 & 1.140 & 0.961 & 0.794 & 0.382 & 0.794 & 0.485 & 0.815\\
\hline
\end{tabular}
\end{center}
\label{conclusion}
\end{table*}

C) As the necessary conditions were satisfied, we used the Gaia DR3 parallax to estimate the absolute parameters. Based on the calculations, the paths of the primary and secondary stars are treated independently, yielding $a_1(R_{\odot})$ and $a_2(R_{\odot})$, which are expected to be equal or close values. The absolute parameter estimations for the target systems show that the difference between $a_1(R_{\odot})$ and $a_2(R_{\odot})$ was less than about 0.1 (Table \ref{conclusion}), which confirms the suitability of the method. This also serves as evidence of the accuracy of the light curve analysis and the input parameters (\citealt{2024NewA..11002227P}, \citealt{2025MNRAS.538.1427P}).

D) The fillout factor is a parameter that describes the degree of contact between components in close binary systems. Contact binary systems are categorized according to their fillout factor into three classes: deep ($f \geq 50\%$), medium ($25\% \leq f < 50\%$), and shallow ($f < 25\%$) systems \citep{2022AJ....164..202L}. Therefore, based on the light curve solutions, three systems were classified as shallow, four as medium, and one as deep fillout factors (Table \ref{conclusion}).

\begin{figure*}
\centering
\includegraphics[width=0.47\textwidth]{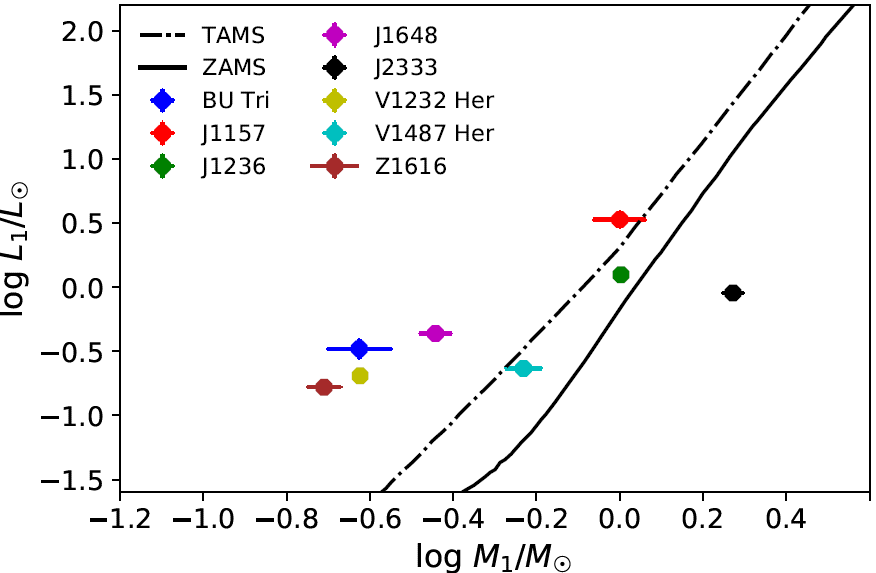}
\includegraphics[width=0.47\textwidth]{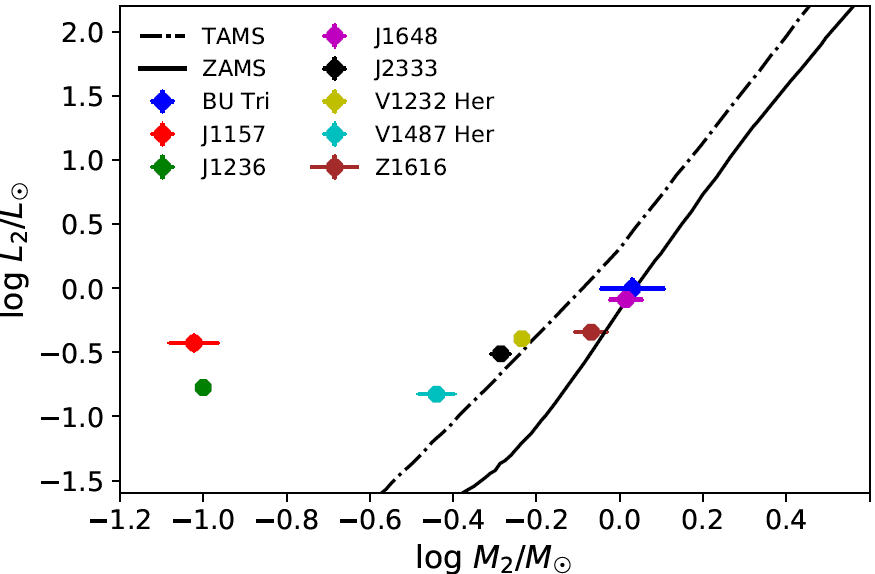}
\includegraphics[width=0.47\textwidth]{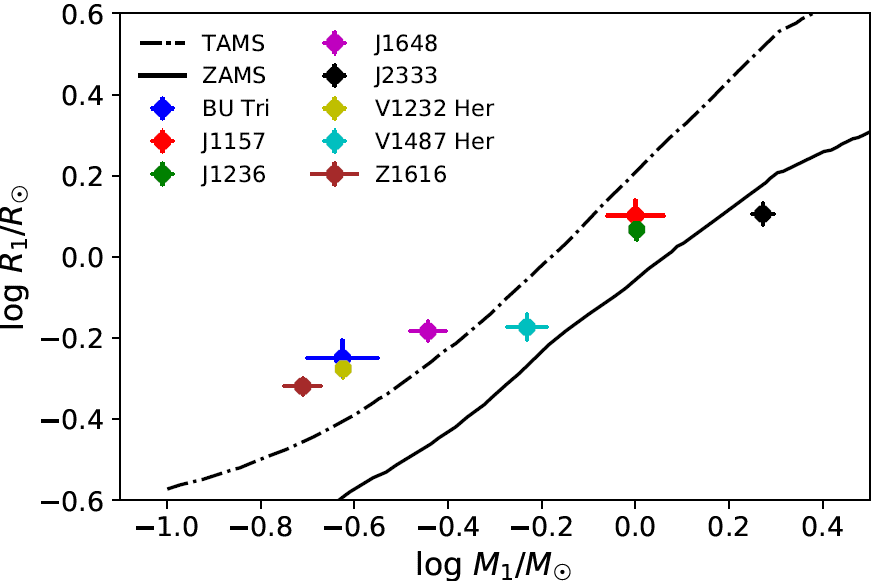}
\includegraphics[width=0.47\textwidth]{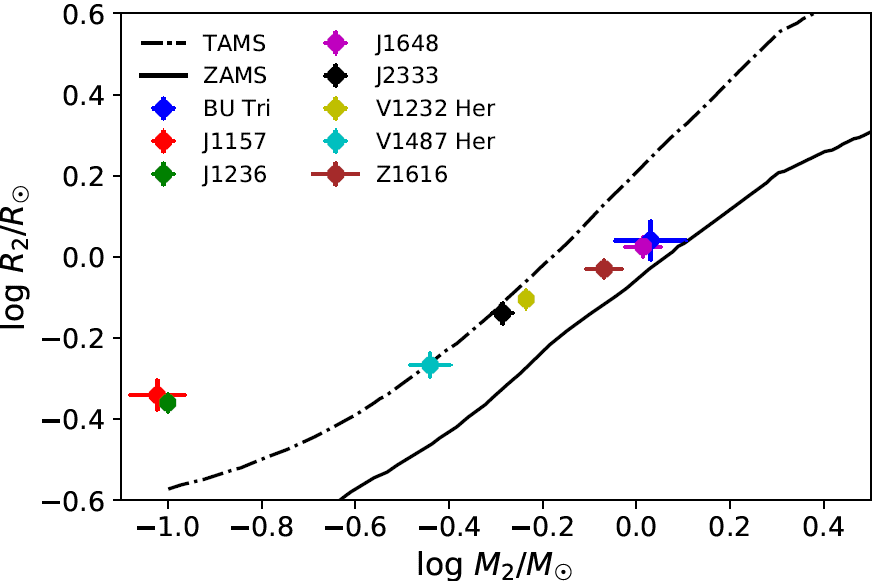}
\caption{Position of both stars in each target system on the $M$–$R$ and $M$–$L$ diagrams.}
\label{MLR}
\end{figure*}

E) The evolutionary status of the systems are presented using logarithmic Mass–Radius ($M$–$R$) and Mass–Luminosity ($M$–$L$) diagrams, derived from the absolute parameters (Table \ref{absolute}, Figure \ref{MLR}). In these diagrams, the stellar components are plotted relative to the Zero-Age Main Sequence (ZAMS) and Terminal-Age Main Sequence (TAMS) lines, as defined by \cite{2000AAS..141..371G}, providing insight into their positions along the stellar evolutionary path.

The light curve solutions and derived absolute parameters show that, in five systems, the hotter component is the less massive star. Conversely, three systems display the opposite configuration, where the more massive star is also the hotter one. As shown in Figure \ref{MLR}, the lower-mass components are generally located near the TAMS, while the more massive ones lie closer to the ZAMS. However, it should be emphasized that contact binaries are products of binary evolution and interaction processes (\citealt{2005ApJ...629.1055Y, 2011AcA....61..139S}, and their evolutionary tracks differ substantially from those of single stars. Therefore, direct comparisons with single-star ZAMS and TAMS lines should be interpreted with caution.

F) To understand the evolution of stars in a contact binary system, it is crucial to determine the initial mass of each component. We utilized the method from \citet{2013MNRAS.430.2029Y} for these calculations. which is based on the assumption that the mass transfer starts near or after the TAMS phase of the massive component (progenitor of the secondary component) and energy transfer has not been taken into account. The initial mass of the secondary star was computed using the following equations, respectively:

\begin{equation}\label{eqM1i}
M_{2i} = M_2 + \Delta M = M_2 + 2.50(M_L - M_2 - 0.07)^{0.64}
\end{equation}

\begin{equation}\label{eqML}
M_L = \left(\frac{L_2}{1.49}\right)^{1/4.216}
\end{equation}

\begin{equation}\label{eqM2i}
M_{1i} = M_1 - (\Delta M - M_{lost}) = M_1 - \Delta M(1 - \gamma)
\end{equation}

\noindent In these equations, $M_1$ and $M_2$ represent the current stellar masses, while $M_{1i}$ and $M_{2i}$ are the initial masses. The parameter $M_L$ is determined from the mass-luminosity relation (Equation \ref{eqML}), and $\Delta M$ represents the mass transferred between the stars. $M_{lost}$ is the mass ejected from the system, and $\gamma$ is the ratio of the mass lost to the total mass transfer, with $\gamma = 0.664$ as adopted from \citet{2013MNRAS.430.2029Y}. The reciprocal mass ratio ($0 < 1/q < 1$) was used in these calculations. The results, shown in Table \ref{conclusion}, are consistent with previous findings by \citet{2013MNRAS.430.2029Y} and \citet{2014MNRAS.437..185Y}.

H) Numerous contact binary systems with low mass ratios ($q \leqslant 0.25$) have been studied (e.g. \citealt{2022AJ....164..202L}, \citealt{2024AJ....168...50L}, \citealt{2024RAA....24j5002S}), yet many questions remain unresolved. Studying contact binaries with extremely low mass ratios is crucial for understanding both the merging process and the lower limit of mass ratios. According to light curve solutions (Table \ref{lc-analysis}), the systems J1157 and J1236 exhibit extremely low mass ratios of $q=$0.095(5) and $q=$0.099(1), respectively. We examined the dynamical stability of these two target systems. Assessing the dynamical stability of contact binaries necessitates knowledge of the ratio between spin angular momentum ($J_{\mathrm{spin}}$) and orbital angular momentum, as outlined by \citet{hut1980stability}. Accordingly, we employed Equation \ref{eqJspin} from \citet{2015AJ....150...69Y} to calculate the $J_{\mathrm{spin}}/J_0$ ratio for the two target systems:

\begin{equation}\label{eqJspin}
\frac{J_{spin}}{J_{0}}=\frac{1+q}{q}[(k_1r_1)^2+(k_2r_2)^2q]
\end{equation}

\noindent where $k_{1,2}$ is the dimensionless gyration radius, $r_{1,2}$ the relative stellar radius, and $q$ is the mass ratio of the system.

We adopted the values of $k_{1,2}$ from \citet{2006MNRAS.369.2001L}. The resulting $J_{\mathrm{spin}}/J_0$ ratios are 0.248 for J1157 and 0.241 for J1236. According to the criteria outlined by \citet{2006MNRAS.369.2001L}, these values indicate that both systems are dynamically stable.

\vspace{0.6cm}
\section*{Data Availability}
Ground-based data are available in the paper's online supplement.

\vspace{0.6cm}
\section*{Acknowledgments}
This manuscript, including the observation, analysis, and writing processes, was provided by the BSN project (\url{https://bsnp.info/}). Work by Kai Li was supported by the National Natural Science Foundation of China (NSFC) (No. 12273018) and by the Qilu Young Researcher Project of Shandong University. This paper is based on observations carried out at the Observatorio Astron\'omico Nacional on the Sierra San Pedro M\'artir which is operated by the Universidad Nacional Aut\'onoma de M\'exico. We used IRAF, distributed by the National Optical Observatories and operated by the Association of Universities for Research in Astronomy, Inc., under a cooperative agreement with the National Science Foundation. We used data from the European Space Agency mission Gaia (\url{http://www.cosmos.esa.int/gaia}). The authors would like to express their gratitude to Dr. David Valls-Gabaud for all his help and advice.

\vspace{0.6cm}
\section*{ORCID iDs}
\noindent Atila Poro: 0000-0002-0196-9732\\
Kai Li: 0000-0003-3590-335X\\
Raul Michel: 0000-0003-1263-808X\\
Li-heng Wang: 0009-0005-0485-418X\\
Fahri Alicavus: 0000-0002-1972-8400\\
Ghazal Alizadeh: 0009-0005-9482-264X\\
Liliana Altamirano-Devora 0000-0001-7715-2182\\
Francisco Javier Tamayo: 0000-0002-9761-9509\\
Hector Aceves:  0000-0002-7348-8815\\

\vspace{0.6cm}
\bibliography{REFS}{}

\begin{thebibliography}{}
\expandafter\ifx\csname natexlab\endcsname\relax\def\natexlab#1{#1}\fi
\providecommand{\url}[1]{\href{#1}{#1}}
\providecommand{\dodoi}[1]{doi:~\href{http://doi.org/#1}{\nolinkurl{#1}}}
\providecommand{\doeprint}[1]{\href{http://ascl.net/#1}{\nolinkurl{http://ascl.net/#1}}}
\providecommand{\doarXiv}[1]{\href{https://arxiv.org/abs/#1}{\nolinkurl{https://arxiv.org/abs/#1}}}

\bibitem[{{Akerlof} {et~al.}(2000){Akerlof}, {Amrose}, {Balsano}, {Bloch}, {Casperson}, {Fletcher}, {Gisler}, {Hills}, {Kehoe}, {Lee}, {Marshall}, {McKay}, {Pawl}, {Schaefer}, {Szymanski}, \& {Wren}}]{2000AJ....119.1901A}
{Akerlof}, C., {Amrose}, S., {Balsano}, R., {et~al.} 2000, \aj, 119, 1901, \dodoi{10.1086/301321}

\bibitem[{{Alonso} {et~al.}(2004){Alonso}, {Brown}, {Torres}, {Latham}, {Sozzetti}, {Mandushev}, {Belmonte}, {Charbonneau}, {Deeg}, {Dunham}, {O'Donovan}, \& {Stefanik}}]{2004ApJ...613L.153A}
{Alonso}, R., {Brown}, T.~M., {Torres}, G., {et~al.} 2004, \apjl, 613, L153, \dodoi{10.1086/425256}

\bibitem[{{Applegate}(1992)}]{1992ApJ...385..621A}
{Applegate}, J.~H. 1992, \apj, 385, 621, \dodoi{10.1086/170967}

\bibitem[{{Bellm} {et~al.}(2019){Bellm}, {Kulkarni}, {Barlow}, {Feindt}, {Graham}, {Goobar}, {Kupfer}, {Ngeow}, {Nugent}, {Ofek}, {Prince}, {Riddle}, {Walters}, \& {Ye}}]{ztf1}
{Bellm}, E.~C., {Kulkarni}, S.~R., {Barlow}, T., {et~al.} 2019, \pasp, 131, 068003, \dodoi{10.1088/1538-3873/ab0c2a}

\bibitem[{Binnendijk(1970)}]{binnendijk1970}
Binnendijk, L. 1970, vistas in Astronomy, 12, 217

\bibitem[{{Butters} {et~al.}(2010){Butters}, {West}, {Anderson}, {Collier Cameron}, {Clarkson}, {Enoch}, {Haswell}, {Hellier}, {Horne}, {Joshi}, {Kane}, {Lister}, {Maxted}, {Parley}, {Pollacco}, {Smalley}, {Street}, {Todd}, {Wheatley}, \& {Wilson}}]{wasp}
{Butters}, O.~W., {West}, R.~G., {Anderson}, D.~R., {et~al.} 2010, \aap, 520, L10, \dodoi{10.1051/0004-6361/201015655}

\bibitem[{{Castelli} \& {Kurucz}(2004)}]{2004AA...419..725C}
{Castelli}, F., \& {Kurucz}, R. 2004, AAP, 419, 725, \dodoi{10.1051/0004-6361:20040079}

\bibitem[{{Chen} {et~al.}(2020){Chen}, {Wang}, {Deng}, {de Grijs}, {Yang}, \& {Tian}}]{2020ApJS..249...18C}
{Chen}, X., {Wang}, S., {Deng}, L., {et~al.} 2020, \apjs, 249, 18, \dodoi{10.3847/1538-4365/ab9cae}

\bibitem[{{Conroy} {et~al.}(2020){Conroy}, {Kochoska}, {Hey}, \& et~al.}]{2020ApJS..250...34C}
{Conroy}, K.~E., {Kochoska}, A., {Hey}, D., \& et~al. 2020, APJS, 250, 34, \dodoi{10.3847/1538-4365/abb4e2}

\bibitem[{{Cox}(2000)}]{2000asqu.book.....C}
{Cox}, A.~N. 2000, {Allen's astrophysical quantities} (AN Cox ed)

\bibitem[{{Drake} {et~al.}(2014){Drake}, {Graham}, {Djorgovski}, {Catelan}, {Mahabal}, {Torrealba}, {Garc{\'\i}a-{\'A}lvarez}, {Donalek}, {Prieto}, {Williams}, {Larson}, {Christen sen}, {Belokurov}, {Koposov}, {Beshore}, {Boattini}, {Gibbs}, {Hill}, {Kowalski}, {Johnson}, \& {Shelly}}]{2014ApJS..213....9D}
{Drake}, A.~J., {Graham}, M.~J., {Djorgovski}, S.~G., {et~al.} 2014, \apjs, 213, 9, \dodoi{10.1088/0067-0049/213/1/9}

\bibitem[{{Eastman} {et~al.}(2010){Eastman}, {Siverd}, \& {Gaudi}}]{Eastman2010}
{Eastman}, J., {Siverd}, R., \& {Gaudi}, B.~S. 2010, \pasp, 122, 935, \dodoi{10.1086/655938}

\bibitem[{{Eker} {et~al.}(2006){Eker}, {Demircan}, {Bilir}, \& {Karata{\c{s}}}}]{2006MNRAS.373.1483E}
{Eker}, Z., {Demircan}, O., {Bilir}, S., \& {Karata{\c{s}}}, Y. 2006, \mnras, 373, 1483, \dodoi{10.1111/j.1365-2966.2006.11073.x}

\bibitem[{{Eker} {et~al.}(2018){Eker}, {Bak{\i}{\c{s}}}, {Bilir}, {Soydugan}, {Steer}, {Soydugan}, {Bak{\i}{\c{s}}}, {Ali{\c{c}}avu{\c{s}}}, {Aslan}, \& {Alpsoy}}]{2018MNRAS.479.5491E}
{Eker}, Z., {Bak{\i}{\c{s}}}, V., {Bilir}, S., {et~al.} 2018, \mnras, 479, 5491, \dodoi{10.1093/mnras/sty1834}

\bibitem[{{Fabry} {et~al.}(2023){Fabry}, {Marchant}, {Langer}, \& {Sana}}]{2023AA...672A.175F}
{Fabry}, M., {Marchant}, P., {Langer}, N., \& {Sana}, H. 2023, \aap, 672, A175, \dodoi{10.1051/0004-6361/202346277}

\bibitem[{{Flower}(1996)}]{1996ApJ...469..355F}
{Flower}, P.~J. 1996, \apj, 469, 355, \dodoi{10.1086/177785}

\bibitem[{{Gaia Collaboration} {et~al.}(2023){Gaia Collaboration}, {Montegriffo}, {Bellazzini}, {De Angeli}, {Andrae}, {Barstow}, {Bossini}, {Bragaglia}, {Burgess}, {Cacciari}, {Carrasco}, {Chornay}, {Delchambre}, {Evans}, {Fouesneau}, {Fr{\'e}mat}, {Garabato}, {Jordi}, {Manteiga}, {Massari}, {Palaversa}, {Pancino}, {Riello}, {Ruz Mieres}, {Sanna}, {Santove{\~n}a}, {Sordo}, {Vallenari}, {Walton}, {Brown}, {Prusti}, {de Bruijne}, {Arenou}, {Babusiaux}, {Biermann}, {Creevey}, {Ducourant}, {Eyer}, {Guerra}, {Hutton}, {Klioner}, {Lammers}, {Lindegren}, {Luri}, {Mignard}, {Panem}, {Pourbaix}, {Randich}, {Sartoretti}, {Soubiran}, {Tanga}, {Bailer-Jones}, {Bastian}, {Drimmel}, {Jansen}, {Katz}, {Lattanzi}, {van Leeuwen}, {Bakker}, {Casta{\~n}eda}, {Fabricius}, {Galluccio}, {Guerrier}, {Heiter}, {Masana}, {Messineo}, {Mowlavi}, {Nicolas}, {Nienartowicz}, {Pailler}, {Panuzzo}, {Riclet}, {Roux}, {Seabroke}, {Th{\'e}venin}, {Gracia-Abril}, {Portell}, {Teyssier}, {Altmann}, {Audard}, {Bellas-Velidis}, {Benson},
  {Berthier}, {Blomme}, {Busonero}, {Busso}, {C{\'a}novas}, {Carry}, {Cellino}, {Cheek}, {Clementini}, {Damerdji}, {Davidson}, {de Teodoro}, {Nu{\~n}ez Campos}, {Dell'Oro}, {Esquej}, {Fern{\'a}ndez-Hern{\'a}ndez}, {Fraile}, {Garc{\'\i}a-Lario}, {Gosset}, {Haigron}, {Halbwachs}, {Hambly}, {Harrison}, {Hern{\'a}ndez}, {Hestroffer}, {Hodgkin}, {Holl}, {Jan{\ss}en}, {Jevardat de Fombelle}, {Jordan}, {Krone-Martins}, {Lanzafame}, {L{\"o}ffler}, {Marchal}, {Marrese}, {Moitinho}, {Muinonen}, {Osborne}, {Pauwels}, {Recio-Blanco}, {Reyl{\'e}}, {Rimoldini}, {Roegiers}, {Rybizki}, {Sarro}, {Siopis}, {Smith}, {Sozzetti}, {Utrilla}, {van Leeuwen}, {Abbas}, {{\'A}brah{\'a}m}, {Abreu Aramburu}, {Aerts}, {Aguado}, {Ajaj}, {Aldea-Montero}, {Altavilla}, {{\'A}lvarez}, {Alves}, {Anderson}, {Anglada Varela}, {Antoja}, {Baines}, {Baker}, {Balaguer-N{\'u}{\~n}ez}, {Balbinot}, {Balog}, {Barache}, {Barbato}, {Barros}, {Bartolom{\'e}}, {Bassilana}, {Bauchet}, {Becciani}, {Berihuete}, {Bernet}, {Bertone}, {Bianchi}, {Binnenfeld},
  {Blanco-Cuaresma}, {Boch}, {Bombrun}, {Bouquillon}, {Bramante}, {Breedt}, {Bressan}, {Brouillet}, {Brugaletta}, {Bucciarelli}, {Burlacu}, {Butkevich}, {Buzzi}, {Caffau}, {Cancelliere}, {Cantat-Gaudin}, {Carballo}, {Carlucci}, {Carnerero}, {Casamiquela}, {Castellani}, {Castro-Ginard}, {Chaoul}, {Charlot}, {Chemin}, {Chiaramida}, {Chiavassa}, {Comoretto}, {Contursi}, {Cooper}, {Cornez}, {Cowell}, {Crifo}, {Cropper}, {Crosta}, {Crowley}, {Dafonte}, \& {Dapergolas}}]{2023AA...674A..33G}
{Gaia Collaboration}, {Montegriffo}, P., {Bellazzini}, M., {et~al.} 2023, \aap, 674, A33, \dodoi{10.1051/0004-6361/202243709}

\bibitem[{{Girardi} {et~al.}(2000){Girardi}, {Bressan}, {Bertelli}, \& {Chiosi}}]{2000AAS..141..371G}
{Girardi}, L., {Bressan}, A., {Bertelli}, G., \& {Chiosi}, C. 2000, AAPs, 141, 371, \dodoi{10.1051/aas:2000126}

\bibitem[{{Green} {et~al.}(2019){Green}, {Schlafly}, {Zucker}, {Speagle}, \& {Finkbeiner}}]{2019ApJ...887...93G}
{Green}, G.~M., {Schlafly}, E., {Zucker}, C., {Speagle}, J.~S., \& {Finkbeiner}, D. 2019, \apj, 887, 93, \dodoi{10.3847/1538-4357/ab5362}

\bibitem[{{Heinze} {et~al.}(2018){Heinze}, {Tonry}, {Denneau}, {Flewelling}, {Stalder}, {Rest}, {Smith}, {Smartt}, \& {Weiland}}]{2018AJ....156..241H}
{Heinze}, A.~N., {Tonry}, J.~L., {Denneau}, L., {et~al.} 2018, \aj, 156, 241, \dodoi{10.3847/1538-3881/aae47f}

\bibitem[{Hut(1980)}]{hut1980stability}
Hut, P. 1980, Astronomy and Astrophysics, vol. 92, no. 1-2, Dec. 1980, p. 167-170., 92, 167

\bibitem[{{Jayasinghe} {et~al.}(2018){Jayasinghe}, {Kochanek}, {Stanek}, {Shappee}, {Holoien}, {Thompson}, {Prieto}, {Dong}, {Pawlak}, {Shields}, {Pojmanski}, {Otero}, {Britt}, \& {Will}}]{jayasinghe2018}
{Jayasinghe}, T., {Kochanek}, C.~S., {Stanek}, K.~Z., {et~al.} 2018, \mnras, 477, 3145, \dodoi{10.1093/mnras/sty838}

\bibitem[{{Kopal}(1959)}]{1959cbs..book.....K}
{Kopal}, Z. 1959, {Close binary systems} (The International Astrophysics Series, London: Chapman \& Hall)

\bibitem[{{Kouzuma}(2023)}]{2023ApJ...958...84K}
{Kouzuma}, S. 2023, \apj, 958, 84, \dodoi{10.3847/1538-4357/ad03e1}

\bibitem[{{Kummer} {et~al.}(2023){Kummer}, {Toonen}, \& {de Koter}}]{2023AA...678A..60K}
{Kummer}, F., {Toonen}, S., \& {de Koter}, A. 2023, \aap, 678, A60, \dodoi{10.1051/0004-6361/202347179}

\bibitem[{{Kwee}(1958)}]{k1958}
{Kwee}, K.~K. 1958, \bain, 14, 131

\bibitem[{{Kwee} \& {van Woerden}(1956)}]{kw1956}
{Kwee}, K.~K., \& {van Woerden}, H. 1956, \bain, 12, 327

\bibitem[{{Lalounta} {et~al.}(2024){Lalounta}, {Christopoulou}, {Papageorgiou}, {Ferreira Lopes}, \& {Catelan}}]{2024AJ....168...50L}
{Lalounta}, E., {Christopoulou}, P.-E., {Papageorgiou}, A., {Ferreira Lopes}, C.~E., \& {Catelan}, M. 2024, \aj, 168, 50, \dodoi{10.3847/1538-3881/ad4882}

\bibitem[{{Lanza} \& {Rodon{\`o}}(1999)}]{1999A&A...349..887L}
{Lanza}, A.~F., \& {Rodon{\`o}}, M. 1999, \aap, 349, 887

\bibitem[{{Latkovi{\'c}} {et~al.}(2021){Latkovi{\'c}}, {{\v{C}}eki}, \& {Lazarevi{\'c}}}]{2021ApJS..254...10L}
{Latkovi{\'c}}, O., {{\v{C}}eki}, A., \& {Lazarevi{\'c}}, S. 2021, \apjs, 254, 10, \dodoi{10.3847/1538-4365/abeb23}

\bibitem[{{Lehky} \& {Horalek}(2007)}]{2007OEJV...58....1L}
{Lehky}, M., \& {Horalek}, P. 2007, Open European Journal on Variable Stars, 0058, 1

\bibitem[{{Li} {et~al.}(2022){Li}, {Gao}, {Liu}, {Gao}, {Li}, {Chen}, \& {Sun}}]{2022AJ....164..202L}
{Li}, K., {Gao}, X., {Liu}, X.-Y., {et~al.} 2022, \aj, 164, 202, \dodoi{10.3847/1538-3881/ac8ff2}

\bibitem[{Li {et~al.}(2020)Li, Kim, Xia, Michel, Hu, Gao, Guo, \& Chen}]{Li_2020}
Li, K., Kim, C.-H., Xia, Q.-Q., {et~al.} 2020, The Astronomical Journal, 159, 189, \dodoi{10.3847/1538-3881/ab7cda}

\bibitem[{{Li} \& {Zhang}(2006)}]{2006MNRAS.369.2001L}
{Li}, L., \& {Zhang}, F. 2006, \mnras, 369, 2001, \dodoi{10.1111/j.1365-2966.2006.10462.x}

\bibitem[{{Liu} \& {Yang}(2003)}]{2003ChJAA...3..142L}
{Liu}, Q.-Y., \& {Yang}, Y.-L. 2003, \cjaa, 3, 142, \dodoi{10.1088/1009-9271/3/2/142}

\bibitem[{{Lucy}(1967)}]{1967ZA.....65...89L}
{Lucy}, L. 1967, ZAP, 65, 89

\bibitem[{{Lucy}(1968{\natexlab{a}})}]{1968ApJ...151.1123L}
{Lucy}, L.~B. 1968{\natexlab{a}}, \apj, 151, 1123, \dodoi{10.1086/149510}

\bibitem[{{Lucy}(1968{\natexlab{b}})}]{1968ApJ...153..877L}
---. 1968{\natexlab{b}}, \apj, 153, 877, \dodoi{10.1086/149712}

\bibitem[{{Masci} {et~al.}(2019){Masci}, {Laher}, {Rusholme}, {Shupe}, {Groom}, {Surace}, {Jackson}, {Monkewitz}, {Beck}, {Flynn}, {Terek}, {Landry}, {Hacopians}, {Desai}, {Howell}, {Brooke}, {Imel}, {Wachter}, {Ye}, {Lin}, {Cenko}, {Cunningham}, {Rebbapragada}, {Bue}, {Miller}, {Mahabal}, {Bellm}, {Patterson}, {Juri{\'c}}, {Golkhou}, {Ofek}, {Walters}, {Graham}, {Kasliwal}, {Dekany}, {Kupfer}, {Burdge}, {Cannella}, {Barlow}, {Van Sistine}, {Giomi}, {Fremling}, {Blagorodnova}, {Levitan}, {Riddle}, {Smith}, {Helou}, {Prince}, \& {Kulkarni}}]{ztf2}
{Masci}, F.~J., {Laher}, R.~R., {Rusholme}, B., {et~al.} 2019, \pasp, 131, 018003, \dodoi{10.1088/1538-3873/aae8ac}

\bibitem[{{O'Connell}(1951)}]{1951PRCO....2...85O}
{O'Connell}, D.~J.~K. 1951, Publications of the Riverview College Observatory, 2, 85

\bibitem[{Paki {et~al.}(2025)Paki, Poro, \& Moosavi~Rowzati}]{paki2025bsn}
Paki, E., Poro, A., \& Moosavi~Rowzati, M.~D. 2025, Galaxies, 13, 74

\bibitem[{{Poro} {et~al.}(2025{\natexlab{a}}){Poro}, {Jahangiri}, {Sarvari}, {Aliakbari}, {Olya}, {Michel}, \& {Tanriver}}]{2025MNRAS.538.1427P}
{Poro}, A., {Jahangiri}, E., {Sarvari}, E., {et~al.} 2025{\natexlab{a}}, \mnras, 538, 1427, \dodoi{10.1093/mnras/staf356}

\bibitem[{{Poro} {et~al.}(2024{\natexlab{a}}){Poro}, {Tanriver}, {Michel}, \& {Paki}}]{2024PASP..136b4201P}
{Poro}, A., {Tanriver}, M., {Michel}, R., \& {Paki}, E. 2024{\natexlab{a}}, \pasp, 136, 024201, \dodoi{10.1088/1538-3873/ad1ed3}

\bibitem[{{Poro} {et~al.}(2024{\natexlab{b}}){Poro}, {Paki}, {Alizadehsabegh}, {Khodadadilori}, {Salehian}, {Hedayatjoo}, {Hashemi}, {Dashti}, \& {Mohammadizadeh}}]{2024RAA....24a5002P}
{Poro}, A., {Paki}, E., {Alizadehsabegh}, A., {et~al.} 2024{\natexlab{b}}, Research in Astronomy and Astrophysics, 24, 015002, \dodoi{10.1088/1674-4527/ad0866}

\bibitem[{{Poro} {et~al.}(2024{\natexlab{c}}){Poro}, {Tanriver}, {Sarvari}, {Zavvarei}, {Azarara}, {Corp}, {Baudart}, {Ababafi}, {Kahali Poor}, {Zare}, {Bulut}, \& {Keskin}}]{2024RAA....24d5018P}
{Poro}, A., {Tanriver}, M., {Sarvari}, E., {et~al.} 2024{\natexlab{c}}, Research in Astronomy and Astrophysics, 24, 045018, \dodoi{10.1088/1674-4527/ad30b2}

\bibitem[{{Poro} {et~al.}(2024{\natexlab{d}}){Poro}, {Li}, {Michel}, {Castro}, {Fern{\'a}ndez Laj{\'u}s}, {Wang}, {Coliac}, {Alada{\u{g}}}, {Alizadehsabegh}, \& {Alicavus}}]{2024AJ....168..272P}
{Poro}, A., {Li}, K., {Michel}, R., {et~al.} 2024{\natexlab{d}}, \aj, 168, 272, \dodoi{10.3847/1538-3881/ad8345}

\bibitem[{{Poro} {et~al.}(2024{\natexlab{e}}){Poro}, {Hedayatjoo}, {Nastaran}, {Nourmohammad}, {Azarara}, {AlipourSoudmand}, {AzarinBarzandig}, {Aliakbari}, {Nasirian}, \& {Kahali Poor}}]{2024NewA..11002227P}
{Poro}, A., {Hedayatjoo}, M., {Nastaran}, M., {et~al.} 2024{\natexlab{e}}, \na, 110, 102227, \dodoi{10.1016/j.newast.2024.102227}

\bibitem[{{Poro} {et~al.}(2025{\natexlab{b}}){Poro}, {Li}, {Paki}, {Baudart}, {Michel}, {Wang}, {Fern{\'a}ndez Laj{\'u}s}, {Alicavus}, {Foschino}, {Aceves}, {Tamayo}, \& {Chavez}}]{2025MNRAS.537.3160P}
{Poro}, A., {Li}, K., {Paki}, E., {et~al.} 2025{\natexlab{b}}, \mnras, 537, 3160, \dodoi{10.1093/mnras/staf222}

\bibitem[{{Poro} {et~al.}(2025{\natexlab{c}}){Poro}, {Li}, , {Michel}, {Wang}, {Alicavus}, {Rhai Nájera}, {Santill\'an-Ortega}, {Tamayo}, \& {Aceves}}]{2025.AJ.P}
{Poro}, A., {Li}, K., , {et~al.} 2025{\natexlab{c}}, \aj, Under review

\bibitem[{{Pr{\v{s}}a} {et~al.}(2016){Pr{\v{s}}a}, {Conroy}, {Horvat}, {Pablo}, {Kochoska}, {Bloemen}, {Giammarco}, {Hambleton}, \& {Degroote}}]{2016ApJS..227...29P}
{Pr{\v{s}}a}, A., {Conroy}, K.~E., {Horvat}, M., {et~al.} 2016, \apjs, 227, 29, \dodoi{10.3847/1538-4365/227/2/29}

\bibitem[{{Qian} {et~al.}(2014){Qian}, {Wang}, {Zhu}, {Snoonthornthum}, {Wang}, {Zhao}, {Zhou}, {Liao}, \& {Liu}}]{2014ApJS..212....4Q}
{Qian}, S.~B., {Wang}, J.~J., {Zhu}, L.~Y., {et~al.} 2014, \apjs, 212, 4, \dodoi{10.1088/0067-0049/212/1/4}

\bibitem[{{Ricker} {et~al.}(2015){Ricker}, {Winn}, {Vanderspek}, {Latham}, {Bakos}, {Bean}, {Berta-Thompson}, {Brown}, {Buchhave}, {Butler}, {Butler}, {Chaplin}, {Charbonneau}, {Christensen-Dalsgaard}, {Clampin}, {Deming}, {Doty}, {De Lee}, {Dressing}, {Dunham}, {Endl}, {Fressin}, {Ge}, {Henning}, {Holman}, {Howard}, {Ida}, {Jenkins}, {Jernigan}, {Johnson}, {Kaltenegger}, {Kawai}, {Kjeldsen}, {Laughlin}, {Levine}, {Lin}, {Lissauer}, {MacQueen}, {Marcy}, {McCullough}, {Morton}, {Narita}, {Paegert}, {Palle}, {Pepe}, {Pepper}, {Quirrenbach}, {Rinehart}, {Sasselov}, {Sato}, {Seager}, {Sozzetti}, {Stassun}, {Sullivan}, {Szentgyorgyi}, {Torres}, {Udry}, \& {Villasenor}}]{tess}
{Ricker}, G.~R., {Winn}, J.~N., {Vanderspek}, R., {et~al.} 2015, Journal of Astronomical Telescopes, Instruments, and Systems, 1, 014003, \dodoi{10.1117/1.JATIS.1.1.014003}

\bibitem[{{Ruci{\'n}ski}(1969)}]{1969AcA....19..245R}
{Ruci{\'n}ski}, S. 1969, ACTAA, 19, 245

\bibitem[{{S{\'a}nchez-S{\'a}ez} {et~al.}(2023){S{\'a}nchez-S{\'a}ez}, {Arredondo}, {Bayo}, {Ar{\'e}valo}, {Bauer}, {Cabrera-Vives}, {Catelan}, {Coppi}, {Est{\'e}vez}, {F{\"o}rster}, {Hern{\'a}ndez-Garc{\'\i}a}, {Huijse}, {Kurtev}, {Lira}, {Mu{\~n}oz Arancibia}, \& {Pignata}}]{2023AA...675A.195S}
{S{\'a}nchez-S{\'a}ez}, P., {Arredondo}, J., {Bayo}, A., {et~al.} 2023, \aap, 675, A195, \dodoi{10.1051/0004-6361/202346077}

\bibitem[{{Sarvari} {et~al.}(2024){Sarvari}, {Fern{\'a}ndez Laj{\'u}s}, \& {Poro}}]{2024RAA....24j5002S}
{Sarvari}, E., {Fern{\'a}ndez Laj{\'u}s}, E., \& {Poro}, A. 2024, Research in Astronomy and Astrophysics, 24, 105002, \dodoi{10.1088/1674-4527/ad7793}

\bibitem[{{Shappee} {et~al.}(2014){Shappee}, {Prieto}, {Grupe}, {Kochanek}, {Stanek}, {De Rosa}, {Mathur}, {Zu}, {Peterson}, {Pogge}, {Komossa}, {Im}, {Jencson}, {Holoien}, {Basu}, {Beacom}, {Szczygie{\l}}, {Brimacombe}, {Adams}, {Campillay}, {Choi}, {Contreras}, {Dietrich}, {Dubberley}, {Elphick}, {Foale}, {Giustini}, {Gonzalez}, {Hawkins}, {Howell}, {Hsiao}, {Koss}, {Leighly}, {Morrell}, {Mudd}, {Mullins}, {Nugent}, {Parrent}, {Phillips}, {Pojmanski}, {Rosing}, {Ross}, {Sand}, {Terndrup}, {Valenti}, {Walker}, \& {Yoon}}]{shappee2014}
{Shappee}, B.~J., {Prieto}, J.~L., {Grupe}, D., {et~al.} 2014, \apj, 788, 48, \dodoi{10.1088/0004-637X/788/1/48}

\bibitem[{{Soomandar} \& {Poro}(2024)}]{2024NewA..10502112S}
{Soomandar}, S., \& {Poro}, A. 2024, \na, 105, 102112, \dodoi{10.1016/j.newast.2023.102112}

\bibitem[{{Sriram} {et~al.}(2017){Sriram}, {Malu}, {Choi}, \& {Vivekananda Rao}}]{2017AJ....153..231S}
{Sriram}, K., {Malu}, S., {Choi}, C.~S., \& {Vivekananda Rao}, P. 2017, \aj, 153, 231, \dodoi{10.3847/1538-3881/aa6893}

\bibitem[{{Stepien}(2011)}]{2011AcA....61..139S}
{Stepien}, K. 2011, \actaa, 61, 139

\bibitem[{{Terrell} \& {Wilson}(2005)}]{2005ApSS.296..221T}
{Terrell}, D., \& {Wilson}, R.~E. 2005, \apss, 296, 221, \dodoi{10.1007/s10509-005-4449-4}

\bibitem[{{Tody}(1986)}]{1986SPIE..627..733T}
{Tody}, D. 1986, in Society of Photo-Optical Instrumentation Engineers (SPIE) Conference Series, Vol. 627, Instrumentation in astronomy VI, ed. D.~L. {Crawford}, 733, \dodoi{10.1117/12.968154}

\bibitem[{{Torres}(2010)}]{2010AJ....140.1158T}
{Torres}, G. 2010, \aj, 140, 1158, \dodoi{10.1088/0004-6256/140/5/1158}

\bibitem[{{Wadhwa} {et~al.}(2024){Wadhwa}, {Landin}, {Kosti{\'c}}, {Vince}, {Arbutina}, {De Horta}, {Filipovi{\'c}}, {Tothill}, {Petrovi{\'c}}, \& {Djura{\v{s}}evi{\'c}}}]{2024MNRAS.527....1W}
{Wadhwa}, S.~S., {Landin}, N.~R., {Kosti{\'c}}, P., {et~al.} 2024, \mnras, 527, 1, \dodoi{10.1093/mnras/stad3129}

\bibitem[{{Yakut} \& {Eggleton}(2005)}]{2005ApJ...629.1055Y}
{Yakut}, K., \& {Eggleton}, P.~P. 2005, \apj, 629, 1055, \dodoi{10.1086/431300}

\bibitem[{{Yang} \& {Qian}(2015)}]{2015AJ....150...69Y}
{Yang}, Y.-G., \& {Qian}, S.-B. 2015, \aj, 150, 69, \dodoi{10.1088/0004-6256/150/3/69}

\bibitem[{{Y{\i}ld{\i}z}(2014)}]{2014MNRAS.437..185Y}
{Y{\i}ld{\i}z}, M. 2014, \mnras, 437, 185, \dodoi{10.1093/mnras/stt1874}

\bibitem[{{Yildiz} \& {Do{\u{g}}an}(2013)}]{2013MNRAS.430.2029Y}
{Yildiz}, M., \& {Do{\u{g}}an}, T. 2013, \mnras, 430, 2029, \dodoi{10.1093/mnras/stt028}

\bibitem[{{Zhang} \& {Qian}(2020)}]{2020MNRAS.497.3493Z}
{Zhang}, X.-D., \& {Qian}, S.-B. 2020, \mnras, 497, 3493, \dodoi{10.1093/mnras/staa2166}

\bibitem[{{Zhou} \& {Leung}(1990)}]{1990ApJ...355..271Z}
{Zhou}, D.-Q., \& {Leung}, K.-C. 1990, \apj, 355, 271, \dodoi{10.1086/168760}

\bibitem[{{Zhou} {et~al.}(2016){Zhou}, {Qian}, {Zhang}, {Jiang}, {Zhang}, \& {Kreiner}}]{2016ApJ...817..133Z}
{Zhou}, X., {Qian}, S.~B., {Zhang}, J., {et~al.} 2016, \apj, 817, 133, \dodoi{10.3847/0004-637X/817/2/133}

\end{thebibliography}
\bibliographystyle{aasjournal}

\end{document}